\documentclass[aps, prb, floatfix, twocolumn, notitlepage, superscriptaddress, 10pt]{revtex4-2}

\usepackage{amsmath, amsthm, amssymb, bbold}
\usepackage[normalem]{ulem}
\usepackage{cancel}
\usepackage{bm}
\usepackage{microtype}
\usepackage{setspace}
\usepackage{graphicx}   
\usepackage{xcolor, color}     
\usepackage[normalem]{ulem}
\usepackage{natbib}
\usepackage{fixmath}
\usepackage{hyperref} 

\newcommand{\pmat}[1]{\begin{pmatrix}#1\end{pmatrix}}
\newcommand{\avg}[1]{\langle{}#1\rangle}
\def\e{\varepsilon}
\def\pd{\partial}
\def\mc{\mathcal}
\def\hc{\textrm{H.c.}}
\hypersetup{colorlinks,linkcolor=blue,urlcolor=blue,citecolor=blue}

\newcommand{\ep}[1]{EP\!\textsubscript{$#1$}}

\newcommand{\pt}{$\mathcal{PT}$}
\newcommand{\cT}{$\mathcal{T}$}
\newcommand{\cP}{$\mathcal{P}$}

\begin{document}
\title{Correlations at higher-order exceptional points in non-Hermitian models}
\author{Doru Sticlet}
\email{doru.sticlet@itim-cj.ro}
\affiliation{National Institute for R\&D of Isotopic and Molecular Technologies, 67-103 Donat, 400293 Cluj-Napoca, Romania}
\author{C\u{a}t\u{a}lin Pa\c{s}cu Moca}
\email{mocap@uoradea.ro}
\affiliation{Department of Theoretical Physics, Institute of Physics, Budapest University of Technology and Economics, 	M\H{u}egyetem rkp.~3, H-1111 Budapest, Hungary}
\affiliation{Department  of  Physics,  University  of  Oradea,  410087,  Oradea,  Romania}
\author{Bal\'azs D\'ora}
\email{dora.balazs@ttk.bme.hu}
\affiliation{Department of Theoretical Physics, Institute of Physics, Budapest University of Technology and Economics, M\H{u}egyetem rkp.~3, H-1111 Budapest, Hungary}
\affiliation{MTA-BME Lend\"{u}let Topology and Correlation Research Group, Budapest University of Technology and Economics, M\H{u}egyetem rkp.~3, H-1111 Budapest, Hungary}
\begin{abstract}
We investigate the decay of spatial correlations of \pt-symmetric non-Hermitian one-dimensional models that host higher-order exceptional points.
Beyond a certain correlation length, they develop anomalous power-law behavior that indicates strong suppression of correlations in the non-Hermitian setups as compared to the Hermitian ones. 
The correlation length is also reflected in the entanglement entropy where it marks a change from logarithmic growth at short distance to a constant value at large distance, characteristic of an insulator, despite the spectrum being gapless.  
Two different families of models are investigated, both having a similar spectrum constrained by particle-hole symmetry.
The first  model offers an experimentally attractive way to generate arbitrary higher-order exceptional points and represents a non-Hermitian extension of the Dirac Hamiltonian for general spin.
At the critical point it displays a decay of the correlations $\sim 1/x^2$ and $1/x^3$ irrespective of the order of the exceptional point. 
The second model is constructed using  unidirectional hopping and displays enhanced suppression of correlations
$\sim 1/x^a$, $a\ge 2$ with a power law that depends on the order of the exceptional point. 
\end{abstract}

\maketitle

\section{Introduction}

According to quantum mechanics, a physical system is described by the Hamiltonian, which is typically assumed to be Hermitian. However, it has been understood for some time that non-Hermitian Hamiltonians also provide valuable insights for the dynamics of open quantum systems, as evidenced by numerous studies~\cite{Bender2007,Moiseyev2011,ElGanainy2018,Ashida2020,Bergholtz2021}.
The motivation for this development may be traced back to proposals that explore unique phenomena that challenge traditional quantum mechanics. 
These include the existence of real spectra of non-Hermitian Hamiltonians~\cite{Bender1998,Mostafazadeh2002}, the non-Hermitian skin effect~\cite{Yao2018,Yao2018a}, unidirectional invisibility~\cite{Berry1998}, and novel topological classifications of noninteracting Hamiltonians~\cite{Lieu2018,Kawabata2019,Zhou2019}. 
Initially, the use of classical analogs of the Schr\"odinger equation allowed us  to achieve experimental control over non-Hermitian Hamiltonians in optics and photonics~\cite{Musslimani2008,Makris2008,Lin2011}. 
Only recently, have there been breakthroughs in accessing genuine many-body non-Hermitian Hamiltonians in quantum mechanics~\cite{Choi2010,Klauck2019,Naghiloo2019,Takasu2020}.

This study focuses on the occurrence and consequences of exceptional points (EP) in the energy spectra of non-Hermitian Hamiltonians, which is a phenomenon with no Hermitian equivalent~\cite{Miri2019,Oezdemir2019,Ding2022}. 
At an EP, two or more eigenvalues become degenerate, and the eigenvectors coalesce, such that they no longer form a complete basis to represent the wave function of the system~\cite{Kato1995}. 
The number of vectors $N$ that coalesce determines the order of the EP, with the most common occurrence being $N=2$. 
We will use the notation \ep{N} to refer to an EP of order $N$. The order $N$ affects the system's response when its parameters are adjusted in the vicinity of the non-Hermitian singularity.
Consider, for example, a noninteracting model described by a single-particle non-Hermitian Hamiltonian. If a perturbation of amplitude $\e$ is applied at an \ep{N}, it typically results in a change of the energy splitting on the order of $\sqrt[N]{\e}$, implying that EPs of higher orders have larger energy splitting~\cite{Wiersig2014,Wiersig2020}. This enhancement can be utilized to improve the sensitivity of sensors to small perturbations. However, as noise is also amplified near EPs, practical methods have been proposed to mitigate its effects~\cite{Zhong2019}.

Recent experiments have extensively investigated \ep{2} realizations in various setups~\cite{Lee2009, Guo2009, Liertzer2012, Brandstetter2014, Peng2014}, as there is a rich phenomenology predicted for second order EPs. In these experiments, unique effects such as the interchange of eigenvectors as the system is moved in parameter space around an EP have been tested~\cite{Dembowski2004,Heiss2004,Berry2011,Heiss2012}, with successful results~\cite{Dembowski2001,Gao2015,Doppler2016}.
However, the investigation of higher-order exceptional points (HOEPs) with $N>2$ poses additional challenges, as it is necessary to tune several parameters for their realization. Nevertheless, symmetries play a crucial role in stabilizing HOEPs~\cite{Mandal2021, Delplace2021, Sayyad2022}, and progress has been made in manufacturing them in optical cavities~\cite{Hodaei2017, Chen2017}, optics~\cite{Kaltsas2022, Huang2022,Zuo2022}, optomechanics~\cite{Xu2016,Jing2017,Zhang2018,Xiong2021}, and acoustics~\cite{Ding2016, Fang2021}.
Unencumbered by experimental limitations, theory has extended our understanding of \ep{N}'s in regard to their classifications~\cite{Kawabata2019a}, the dynamics of wave functions around an \ep{N}~\cite{Heiss2008,Demange2011,Hoeller2020}, Landau-Zener tunneling at an EP~\cite{RamyaParkavi2021, Melanathuru2022}, or interactions-induced EPs~\cite{Crippa2021}.

Exceptional points occur naturally in non-Hermitian \pt-symmetric systems~\cite{Bender2019}. These systems have a Hamiltonian that is symmetric under the combined operation of time reversal (\cT) and parity symmetry (\cP).
These systems exhibit two distinct phases: The \pt-symmetric phase where the spectrum of the Hamiltonian is real, and a \pt-broken phase where the spectrum becomes complex~\cite{Rueter2010}. 
The transition between the two phases is marked by the occurrence of the EP. Recently, there has been a growing interest in investigating the properties of the system exactly at the EP~\cite{Ashida2017,Ashida2018,Dora2022}. In the present work, we refer to the state at the exceptional point as the critical state. In particular, Ref.~\cite{Dora2022} demonstrated that the power-law behavior of spatial correlations near an \ep{2} is characterized by anomalous exponents, which signify stronger suppression of correlations in non-Hermitian setups compared to the Hermitian models.

The aim of this  work is two-fold.
First, it seeks to contribute to the quest for natural ways to implement HOEPs in simple lattices, which has been investigated in recent studies~\cite{Zhong2020, Mandal2021, Wiersig2022}.
To do this, we propose a quasi-one-dimensional lattice with gain and loss, which may be used as a model to generate \ep{N} of arbitrary order.
For that we employ a lattice representation of a non-Hermitian Dirac Hamiltonian for general spin $S=(N-1)/2$. 
These systems exhibit a particle-hole symmetry that enforces a flat band at the EP for integer spin $S$ or odd $N$.
The second goal of the study is to investigate the critical state of such systems reflected in the anomalous power-law behavior of spatial correlations and the entanglement entropy.

\begin{figure}[t!]
    \includegraphics[width=\columnwidth]{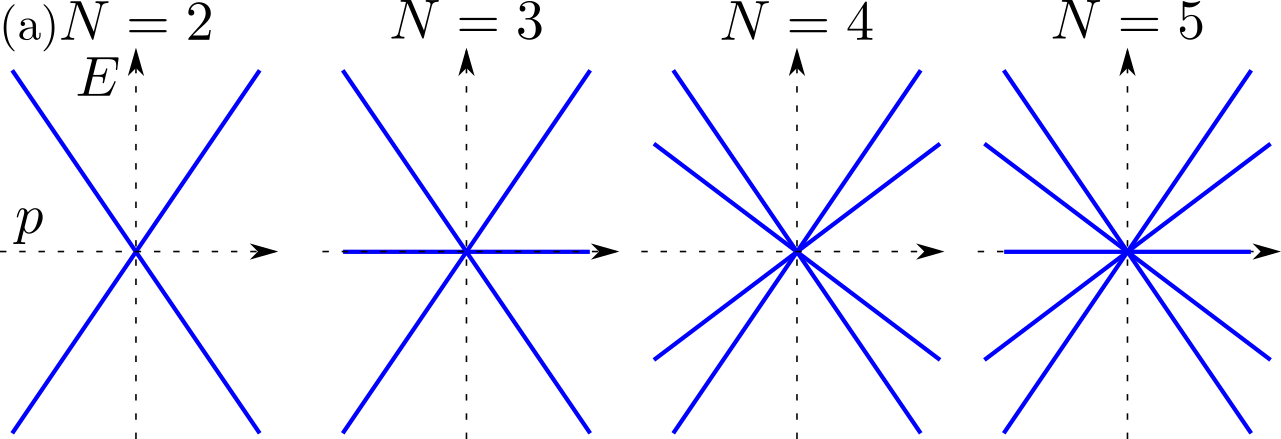}
    \includegraphics[width=\columnwidth]{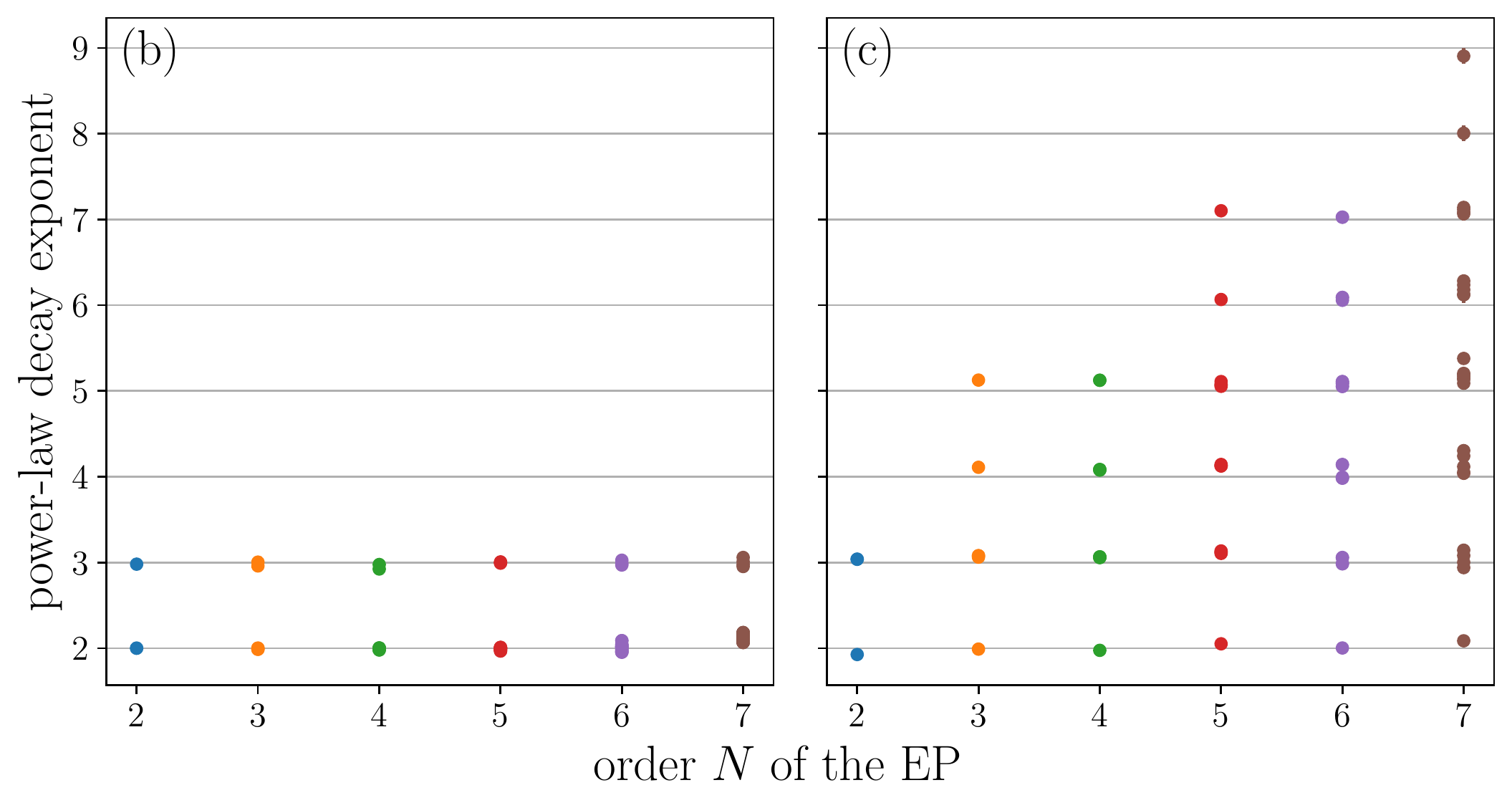}
    \caption{(a) Sketch of energy dispersion (blue lines) in \pt-symmetric models near an \ep{N}. Flat bands occur only for odd $N$. 
    (b, c) The exponents $a$ that describe the algebraic decay $\sim 1/x^a$ for all  $N(N+1)/2$  possible combinations of real-space correlators as a function of the order $N$ of the EP. 
    Exponents are extracted numerically from fitting the large distance $x$ behavior of the corresponding Green's functions in the two classes of the models studied.
    (b) Suppression of correlations is limited to $1/x^2$ and $1/x^3$ decay in the model implementing the non-Hermitian general spin Dirac Hamiltonian irrespective of the order of the EP.
    (c) Correlations display a general $1/x^a$ decay, with $a\ge 2$ that depends on the order of the EP 
    in the unidirectional model.
    }
    \label{fig:power}
\end{figure}

In our study, we thoroughly examine two distinct models, both showcasing HOEPs. The first model is constructed on a diamond lattice, with gain and loss incorporated into the hoppings, and it is associated with a general spin Dirac model. The second model is based on a ladder-like tight-binding structure with unidirectional couplings. Remarkably, our findings reveal that while both models host HOEPs, they exhibit different power-law decay patterns in correlations. The precise algebraic exponents for these models are illustrated in Fig.~\ref{fig:power}, showing in both cases a suppression of correlations with respect to the Hermitian one-dimensional free fermion systems. 
The spin Dirac model displays a limited suppression in the correlation $\sim 1/x^{\alpha}$ 
with $\alpha\in \{2,3\} $, irrespective of the order of the EP, while for the unidirectional model, the exponent increases with the order of the EP.

This paper is structured as follows. In Sec.~\ref{sec:spin}, we present the general spin Dirac model with arbitrary order \ep{N}, and we investigate the correlations and entanglement entropy within it, with a particular focus on the \ep{3} case, which possesses a flat band and is analytically tractable. In section~\ref{sec:unidirectional}, we introduce and examine a second family of models with unidirectional hopping, which exhibit an increasing suppression of correlations with the order of the exceptional point. In Sec.~\ref{sec:comment}, we discuss the relationship between the two families of models in the low-energy limit. The final section, Sec.~\ref{sec:conclusions}, presents the conclusions of our study. 
Additionally, several appendices clarify various points in the main text, such as defining conventions for correlation functions (Sec.~\ref{sec:correlations}), demonstrating the role of particle-hole symmetry in models with a flat band (Sec.~\ref{sec:phs_flat}), or providing examples in computing correlation functions (Sec.~\ref{sec:example}).

\section{Lattice model supporting \texorpdfstring{\ep{N}}{EPN} and the general spin Dirac Hamiltonian}
\label{sec:spin}
In this section, we systematically construct a family of lattice models that host higher-order exceptional points.
These are non-Hermitian models with balanced gain and loss that display a real spectrum due to their \pt-symmetry.

We start with a known one-dimensional tight-binding model introduced before~\cite{Ashida2018,Dora2022} that has a regular \ep{2}, and we demonstrate that combining $N$ such chains produces a quasi-one-dimensional lattice featuring a single \ep{N} point, while maintaining \pt\ symmetry. By tuning the chemical potential and hoppings in the chains, we ensure that the low-energy dispersion is characterized by a non-Hermitian Dirac  Hamiltonian for spin $S=(N-1)/2$. Using models with \ep{N}, we examine the physical implications of the spatial correlation functions that develop.

Our focus is on studying the ground state of the systems at zero temperature, where all energy levels are filled up to the energy associated with the \ep{N}. Due to the real spectrum of the critical non-Hermitian model, the filling of the states is defined in the order of increasing energy. We examine the critical state using correlation functions and find that instead of the typical $\sim 1/x$ decay, characteristic of one-dimensional non-interacting fermions~\cite{Giamarchi2004}, there are anomalous power-law decay laws $\sim 1/x^a$, with $a>1$. Additionally, we observe a significant density-charge imbalance in the model and a persistent ground-state current.

\subsection{Lattice model with \texorpdfstring{\ep{2}}{EP2}}\label{sec:model_EP2}
Let us first discuss the \pt-symmetric tight-binding model hosting an \ep{2}~\cite{Ashida2018,Dora2022},
\begin{equation}\label{h_tb_n2}
H = \sum_{j} (-1)^j\mu c_j^\dag c_j^{}
+ \frac{t-i(-1)^j\gamma}{2}(c_j^\dag c_{j+1}^{}+c_{j+1}^\dag c_j^{}),
\end{equation}
with $c_j^\dag$ ($c_j^{}$) denoting the fermion creation (annihilation) operator at site $j$.
There are two atoms in the unit cell, $a_{n}\equiv c_{2j}$ and $b_{n}\equiv c_{2j+1}$, with cell index $n$. 
There is an alternating gain and loss on neighboring links modeled by the rate $\gamma>0$.
The chemical potential $\mu$ is real, and without loss of generality, the hopping integral is  henceforth considered positive, $t>0$.
In momentum space,
\begin{equation}
H = \sum_k c_k^\dag h_k^{} c_k^{},
\end{equation} 
with $c_k^\dag=(a_k^\dag,b_k^\dag)$. The Bloch Hamiltonian reads,
\begin{equation}\label{bloch_2}
h_k = \pmat{
\mu & f_{k} \\
f_{-k} & -\mu
},
\end{equation}
with 
\begin{equation}\label{fk}
f_k = t\cos(k/2) + \gamma\sin(k/2),
\end{equation}
and $k$ in $(0,2\pi)$. The lattice constant is set throughout to 1 and $\hbar=1$.
This gauge choice of momentum-dependent phases on the hoppings results in a Bloch Hamiltonian that is $4\pi$-periodic, while the dispersion remains $2\pi$-periodic. 
The two energy bands are characterized by the dispersion
\begin{equation}
E_\pm(k)= \pm\sqrt{t^2\cos(k/2)^2+\mu^2-\gamma^2\sin(k/2)^2}.
\end{equation}
Consequently, a single exceptional point develops at $k=\pi$, for $\mu=\pm\gamma$.

An expansion near the EP, $k=\pi+p$, for small momenta $p$, provides an approximation of the model close 
to the \ep{2}, as the effective Hamiltonian becomes
\begin{equation}
h(p)=\pmat{\mu & -pt/2+\gamma \\ -pt/2-\gamma & -\mu}.
\end{equation}
Performing a rotation in the pseudo-spin space, the continuum Hamiltonian near the \ep{2}, e.g., for $\mu=\gamma$, reads,
\begin{equation}\label{h_cont_2}
\tilde h(p)=\pmat{vp & \Delta \\ 0 & -vp},
\end{equation}
with $v=t/2$ and $\Delta=2\gamma$.
This continuum model was analyzed in detail in Ref.~\cite{Dora2022}. Labeling the field operators for right- and left-moving particles as $\psi_1(x)$ and $\psi_2(x)$, respectively, it was found that the correlations behave as
\begin{eqnarray}\label{N2_cont}
    \avg{\psi_1^\dag(x)\psi_1^{}(0)} &=& 
    \frac{i\Delta}{2\pi v}\times
    \begin{cases}
    \frac{v}{\Delta x} & x \ll \frac v\Delta, \notag\\
    - (\frac{2v}{\Delta x})^3 & x\gg \frac v\Delta,
    \end{cases} \\
    \avg{\psi_2^\dag(x)\psi_2^{}(0)} &=& -\avg{\psi_1^\dag(x)\psi_1^{}(0)},\\
    \avg{\psi_1^\dag(x)\psi_2^{}(0)} &=& \frac{\Delta}{4\pi v}\times
    \begin{cases}
    \ln(\frac{\Delta x}{2v}) & x\ll \frac{v}{\Delta},\\
    (\frac{2v}{\Delta x})^2 & x\gg \frac{v}{\Delta},
    \end{cases}\notag
\end{eqnarray}
with propagators computed in the ground state (see Apps.~\ref{sec:correlations} and \ref{sec:phs_flat} for details).
The non-Hermitian term $\Delta$ is not present in the spectrum, but it influences the eigenvectors, and the correlation functions.
Crucially, it introduces a correlation length scale $\xi=v/\Delta$ which marks the crossover from an almost Hermitian regime to the anomalous long-range behavior of correlations.
In the short-distance regime, $x\ll\xi$, the system behaves as in the Hermitian case, $\Delta=0$, ($\avg{\psi_1^\dag(x)\psi_1^{}(0)} = -\avg{\psi_2^\dag(x)\psi_2^{}(0)} \sim i/2\pi x$) with only a logarithmic correction to the off-diagonal propagator $\avg{\psi_1^\dag(x)\psi_2^{}(0)}$, which is otherwise zero in the Hermitian case.
In the large-distance limit, $x\gg\xi$, the correlations decay with an anomalous power law, $1/x^2$ or $1/x^3$. The suppression was interpreted as due to the quantum Zeno effect~\cite{Misra1977,Barontini2013}.

\subsection{Generalization to \texorpdfstring{\ep{N}}{EPN}}

The tight-binding model introduced in Eq.~\eqref{h_tb_n2} consists of a single chain described by a nearest-neighbor hopping with alternating gain and loss, and a chemical potential that depends on the site. 
When connecting $N$ of these one-dimensional chains, each of which supports an \ep{2}, it is possible to create a quasi-one-dimensional structure in the form of a diamond-like lattice that hosts an \ep{N}.
A simple way to understand the concept is by starting with a Bloch Hamiltonian expressed in momentum space, which resembles Eq.~\eqref{bloch_2}, but extending the model from a spin $S=1/2$ representation to a spin of arbitrary value $S$~\cite{Dora2011},
\begin{equation}\label{spin_ham}
h_k = t\cos(k/2)S^x + i\gamma\sin(k/2)S^y + \mu S^z.
\end{equation}
For a larger spin, we construct the $N\times N$ matrices $S^i$ that form a representation for the spin operators $\hat S^i$ using the basis states $|S=\frac12(N-1),m\rangle$ (see details in App.~\ref{sec:spin_mat}). 
The spin quantum number $m$ takes $N$ values, ranging from $-S$ to $S$.
The energy spectrum reads
\begin{equation}\label{e_m}
E_m = m\sqrt{t^2\cos(k/2)^2+\mu^2-\gamma^2\sin(k/2)^2},
\end{equation}
where $m$ is the spin quantum number.
For $|\mu|>\gamma$, the spectrum is real and gapped. For $|\mu|<\gamma$, the spectrum becomes imaginary for some momenta $k$. 
Exactly at $|\mu|=\gamma$, there is a real-to-complex transition marked by an \ep{N}, $N=2S+1$, located at $k=\pi$. In this case, the dispersion simplifies to 
\begin{equation}\label{em_latt}
E_m=m\sqrt{t^2+\gamma^2}|\cos(k/2)|,
\end{equation}
with $N$ bands crossing at the \ep{N}.
Moreover, for integer spin $S$, there are an odd number $N$ of bands including a flat band, corresponding to $m=0$.

\begin{figure}[t]
    \includegraphics[width=\columnwidth]{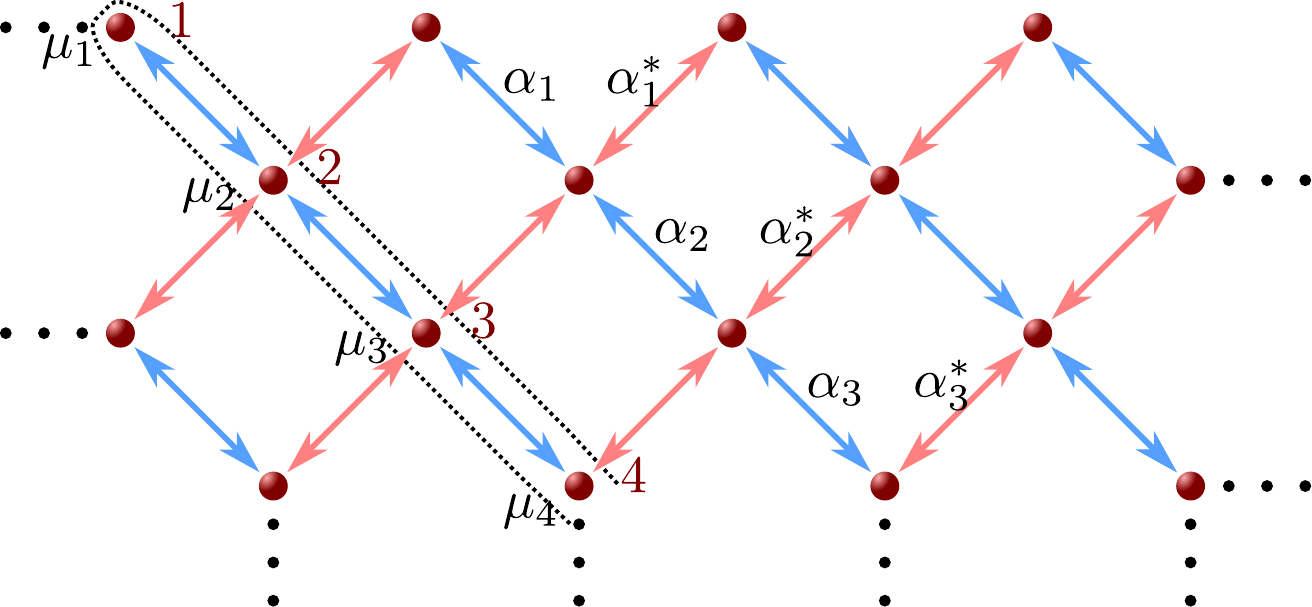}
    \caption{
        The Dirac Hamiltonian with spin $S$ is realized on a diamond lattice that hosts \ep{N}, where $N$ is equal to $2S+1$. Losses and gains in hoppings are denoted by blue and red arrows, respectively, with the coefficients of the hoppings $\alpha_j$ being proportional to $(t-i\gamma)$ for losses and $\alpha_j^*$, to $(t+i\gamma)$ for gains. The unit cell, which is highlighted by a dotted line, contains $N$ atoms, and the sites have chemical potentials $\mu_j$. The amplitudes of the chemical potentials and hoppings $\alpha_j$ in each row $j$ are tuned according to Eq.~\eqref{coeffs}.}
    \label{fig:diamond_latt}
\end{figure}
To investigate the physics of the \ep{N}, we consider the zero-temperature limit and a fixed filling such that the Fermi level corresponds to the energy of the \ep{N}. In particular, in the case of integer $S$, the model contains either an empty or a full flat band (see App.~\ref{sec:phs_flat}).
Similar to the \ep{2} model discussed in Sec.~\ref{sec:model_EP2}, it is fruitful to develop a low-energy description of the model as it encompasses the relevant physics. 
In accordance with the same procedure, the momentum near the EP is expanded as $k=\pi+p$, yielding an effective non-Hermitian continuum model,
\begin{equation}\label{spin_cont_3}
h(p) = -\frac{pt}{2}S^x + i\gamma S^y +\mu S^z.
\end{equation}
When the parameters are set such that $\mu=\pm\gamma$, the system exhibits a linear dispersion near the \ep{N}, $E_m=mpt/2$. 
In Fig.~\ref{fig:power}(a), we display the linear dispersion alongside the formation of flat bands that define the spectrum of Eq.~\eqref{spin_cont_3} for different $N$. 
Essentially, Eq.~\eqref{spin_cont_3} can be regarded as a non-Hermitian Dirac Hamiltonian owing to the linear rates of dissipation.
Moreover, the Hamiltonian at $p=0$ is analyzed to confirm that it describes an \ep{N} with $N$ coalescing eigenvectors [see App.~\ref{sec:epn_proof}].

The Hamiltonian in real space that corresponds to the model~\eqref{spin_ham} is given by
\begin{eqnarray}\label{ham_diamond}
    H &=& \sum_{n=-L}^{L}\bigg[\sum_{j=1}^N\mu^{}_j c^\dag_{j,n} c^{}_{j,n} 
    +\sum_{j=1}^{N-1}\alpha^{}_j(c^\dag_{j,n}c^{}_{j+1,n}+\hc)\bigg]\notag\\   
    &&{}+\sum_{n=-L}^{L-1}\sum_{j=1}^{N-1}\alpha_j^*(c^\dag_{j,n+1}c^{}_{j+1,n}+\hc),
\end{eqnarray}
where
\begin{equation}\label{coeffs}
\mu_j = \mu S^z_{j,j},\quad \alpha_j = \frac{t-i\gamma}{2}S^x_{j,j+1},
\end{equation}
where $\mu$ is a constant chemical potential, and $S^{x,z}$ representing the spin matrices (App.~\ref{sec:spin_mat}).
The creation (annihilation) operators $c^{\dag}_{j,n}$ ($c_{j,n}$) are identified by integers $n$, which represents the unit cell, and $j$, which specifies the site within the unit cell $n$. The Hamiltonian is constructed on a quasi-one-dimensional  diamond lattice, where alternating gain and loss rates $\gamma$ are applied, as shown in Fig.~\ref{fig:diamond_latt}. Throughout this work, we typically consider infinite or periodic lattices, where $L\to\infty$ with open boundary condition or finite L with periodic boundary conditions. 
The width of the stripe in terms of sites is set by the order $N$ of the \ep{N}.

The Hamiltonian in Eq.~\eqref{ham_diamond} exhibits \pt\ symmetry despite the breaking of conventional time-reversal symmetry. When time reversal is applied, the gain and loss rates are interchanged, which corresponds to taking the complex conjugate of Eq.~\eqref{ham_diamond}. 
Since the time-reversal operator for spinless fermions is $\mc T=\mc K$, with $\mc K$ being the complex conjugation operator ($\mc Ki\mc K=-i$), the Hamiltonian is not invariant under $\mc T$. 
However, the application of a reflection operation $\mc P$, which changes $x\to-x$ along the stripe of the diamond lattice, interchanges the rates again, and it makes the Hamiltonian invariant under the joint \pt\ symmetry operations. For instance, by choosing the reflection line to be normal to $x$ and passing through the first atom of cell $n=0$, the effect of $\mc P$ on fermion operators is determined as $\mc P c_{j,n}\mc P^{-1}=c_{j,-n+1-j}$. 
Therefore, the tight-binding Hamiltonian~\eqref{ham_diamond} for $L\to\infty$ becomes indeed invariant under the application of \pt\ symmetry.

\begin{figure}[t]
\includegraphics[width=\columnwidth]{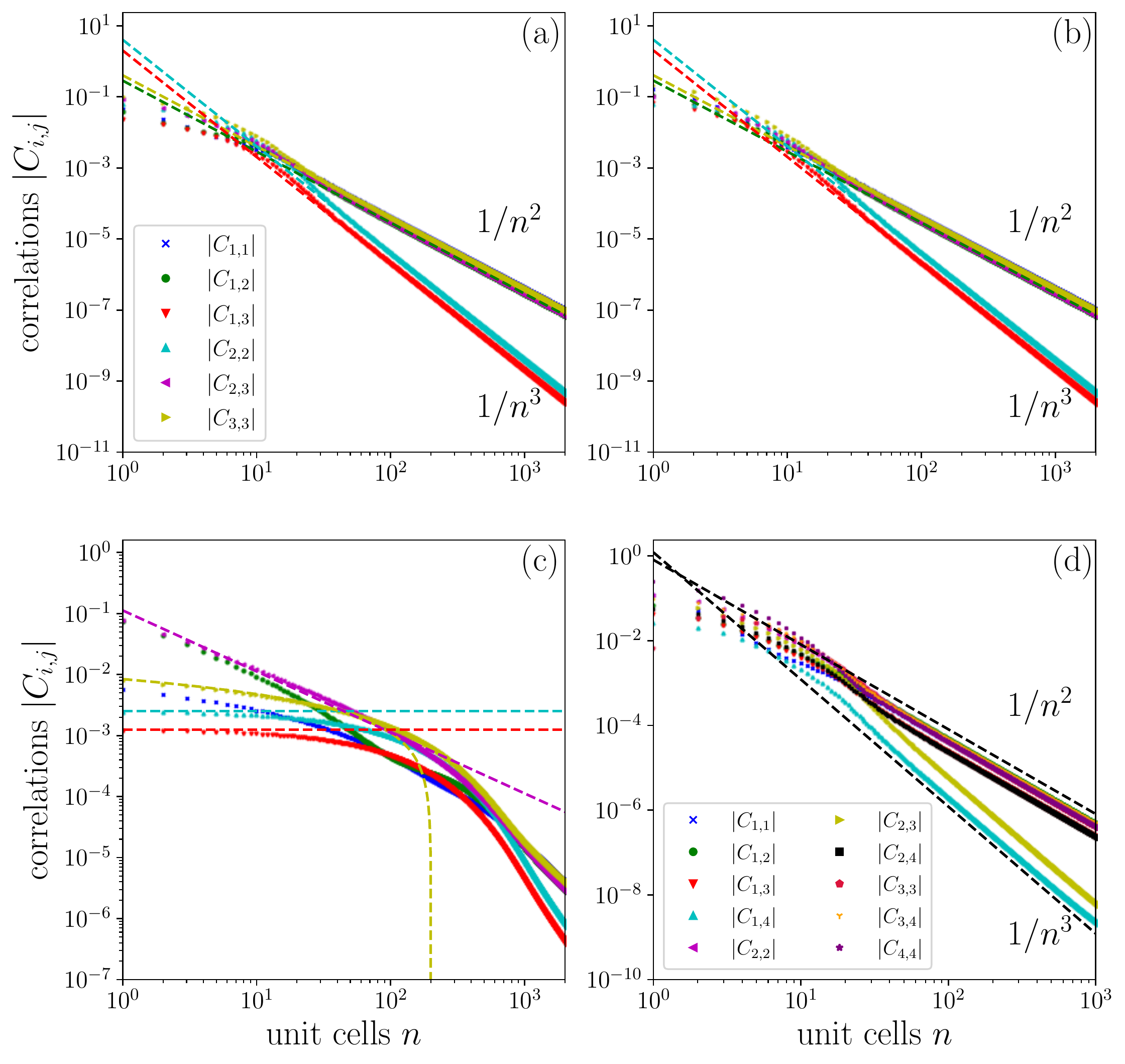}
\caption{Correlation functions in the diamond lattice at EPs, $\mu=\gamma$. 
Correlations for $N=3$, $\gamma=0.2t$ at filling (a) $\nu=1/3$ and (b) $\nu=2/3$ from numerics (colored symbols) follow closely the analytical lattice results (colored dashed lines) at large $n$ from Eqs.~\eqref{corr_diamond_3_diag} and \eqref{corr_diamond_3_off}. (c) Short-distance behavior for $N=3$, $\nu=1/3$, at $\gamma=.005t$ is qualitatively given by the continuum results in Eqs.~\eqref{diamond_cont_intra_3} and \eqref{diamond_cont_inter_3}.
The matching color of a symbol from numerics and the dashed line from analytics identify the same correlation function.
(d) For $N=4$, $\nu=1/2$, and $\gamma=0.2t$, correlations still decay at large distance as $1/n^2$ or $1/n^3$ (the dashed black lines are guidelines).
Panels (a), (b) and (c) share the legend.
}
\label{fig:corr_diamond}
\end{figure}
    
\subsection{Correlations in the \texorpdfstring{\ep{N}}{EPN} models}
\label{sec:diamond_correlations}

In the present sections the main emphasis is on the two-point correlation functions in models that have \ep{N}, with $N>2$. These functions play a crucial role in describing the universal properties at the critical point and in determining the charge density and currents. The case in which $N=3$ is particularly intriguing because it is the simplest one that extends beyond the known results and still admits an analytical solution. What is interesting about this case is that it involves a flat band, and the correlations are computed analytically by assuming the flat band to be either completely empty or filled.

The lattice hosting \ep{3} is realized by connecting two $N=2$ chains shifted by half a lattice constant. 
The  chemical potentials are fixed to  $\mu$, $0$ and $-\mu$. 
Thus, the model consists of only the two upmost rows of hoppings in Fig.~\ref{fig:diamond_latt}.
The alternating hoppings are $(t\pm i\gamma)/2\sqrt2$. 
Consequently, the Bloch Hamiltonian from Eq.~\eqref{spin_ham} explicitly reads
\begin{equation}
h_k = \pmat{
\mu & \frac{1}{\sqrt 2}f_{k} & 0 \\
\frac{1}{\sqrt 2}f_{-k} & 0 & \frac{1}{\sqrt 2}f_k \\
0 & \frac{1}{\sqrt 2}f_{-k} & -\mu
},
\end{equation}
with $f_k$ from Eq.~\eqref{fk}

The spectrum of the Hamiltonian which realizes an \ep{3} for $\mu=\pm\gamma$ consists of three bands crossing at $k=\pi$ with a flat band $E_0=0$ and two dispersing bands
\begin{equation}\label{e_3}
E_\pm = \pm\sqrt{t^2+\gamma^2}|\cos(k/2)|.
\end{equation}

There are six distinct correlation functions 
\begin{equation}
    C_{i,j}(n)=\avg{c^\dag_{i,n} c^{}_{j,0}}, \phantom{aa}i\leq j,\phantom{aa} i,j\in\{1,2,3\}.
\end{equation}
First, we investigate the case in which the filling is $\nu=1/3$ which implies that the flat band is completely empty.
The numerical results are presented in Fig.~\ref{fig:corr_diamond}(a) and (c). 
In the asymptotic large-distance limit $n\gg t/\gamma$, the same correlations are computed analytically by integrating over the occupied eigenstates of the Bloch Hamiltonian (see App.~\ref{sec:example}). 
The diagonal correlations are in the asymptotic limit,
\begin{align}\label{corr_diamond_3_diag}
\avg{c_{1,n}^\dag c^{}_{1,0}} &\sim \frac{\sqrt{t^2+\gamma^2}}{4\pi\gamma}\frac{\cos(\pi n)}{n^2},\notag\\
\avg{c_{2,n}^\dag c^{}_{2,0}} &\sim -\frac{it\sqrt{t^2+\gamma^2}}{2\pi\gamma^2}\frac{\cos(\pi n)}{n^3},\\
\avg{c_{3,n}^\dag c^{}_{3,0}} &\sim -\frac{\sqrt{t^2+\gamma^2}}{4\pi\gamma}\frac{\cos(\pi n)}{n^2},\notag
\end{align}
and the off-diagonal correlations,
\begin{align}\label{corr_diamond_3_off}
\avg{c_{1,n}^\dag c^{}_{2,0}} &\sim -\frac{\sqrt{2t^2+2\gamma^2}}{8\pi\gamma}\frac{\cos(\pi n)}{n^2},\notag\\
\avg{c_{1,n}^\dag c^{}_{3,0}}&\sim -\frac{it\sqrt{t^2+\gamma^2}}{4\pi\gamma^2}\frac{\cos(\pi n)}{n^3},\\
\avg{c_{2,n}^\dag c^{}_{3,0}} &\sim \frac{\sqrt{2t^2+2\gamma^2}}{8\pi\gamma}\frac{\cos(\pi n)}{n^2}.\notag
\end{align}
The numerical results, depicted in Fig.~\ref{fig:corr_diamond}(a), are in excellent agreement with the analytical calculation. The correlations display an anomalous decay pattern similar to that of the $N=2$ case, with the correlation decaying at large distances with $\sim 1/n^2$ and $\sim 1/n^3$.
The same results are obtained when considering a filled flat band with $\nu=2/3$. Exploiting the particle-hole symmetry in the system simplifies the process of determining the correlation functions based on the $\nu=1/3$ results (see App.~\ref{sec:phs_flat}).
\begin{equation}\label{phs_latt_3}
C^{\nu=2/3}_{i,j}(n) = (-1)^{i+j+1}C^{\nu=1/3}_{4-j,4-i}(n),
\end{equation}
where the filling is denoted explicitly in the superscript. 
Numerical results shown in Fig.~\ref{fig:corr_diamond}(b) substantiate this claim and confirm that the results at $\nu=2/3$ are unchanged with respect to the $\nu=1/3$ case, when taking the absolute value of the correlation functions.
The continuum Hamiltonian~\eqref{spin_cont_3} at the exceptional point, e.g.,~$\mu=\gamma$, allows one to compute correlations analytically both in the large-distance and short-distance limits. 
The diagonal correlations read
\begin{align}\label{diamond_cont_intra_3}
    \avg{\psi^\dag_{1}(x) \psi^{}_{1}(0)} &\sim \frac{1}{4\pi\xi}
    \times
    \begin{cases}
    4\ln(x/\xi) &x\ll \xi,\\
    \xi^2/x^2 &x\gg \xi,
    \end{cases}
    \notag\\
    \avg{\psi^\dag_{2}(x) \psi^{}_{2}(0)} &\sim -\frac{1}{2\pi\xi}
    \times
    \begin{cases}
    \pi &x\ll \xi,\\
    i\xi^3/x^3 &x\gg \xi,
    \end{cases}
    \\
    \avg{\psi^\dag_{3}(x) \psi^{}_{3}(0)} &\sim -\frac{1}{4\pi\xi}
    \times
    \begin{cases}
    4\ln(x/\xi) &x\ll \xi,\\
    \xi^2/x^2 &x\gg \xi,
    \end{cases}
    \notag
\end{align}
and the off-diagonal ones,
\begin{align}\label{diamond_cont_inter_3}
    \avg{\psi^\dag_{1}(x) \psi^{}_{2}(0)} &\sim -\frac{\sqrt2}{8\pi\xi}
    \times
    \begin{cases}
    2i\xi/x &x\ll \xi,\\
    \xi^2/x^2 &x\gg \xi,
    \end{cases}
    \notag\\
    \avg{\psi^\dag_{1}(x) \psi^{}_{3}(0)}&\sim -\frac{1}{4\pi\xi}
    \times
    \begin{cases}
    \pi &x\ll \xi,\\
    i\xi^3/x^3 &x\gg \xi,
    \end{cases}
    \\   
    \avg{\psi^\dag_{2}(x) \psi^{}_{3}(0)} &\sim \frac{\sqrt 2}{8\pi\xi}
    \times
    \begin{cases}
    -2i\xi/x &x\ll \xi,\\
    \xi^2/x^2 &x\gg \xi.
    \end{cases}
    \notag
\end{align}
These results indicate that the anomalous decay regime is determined by a natural correlation length that depends on the dissipation rates, $\xi=t/\gamma$.
It should be noted that the continuum Hamiltonian, valid to first order in $p$, fails to replicate the actual Fermi velocity obtained from the lattice model $\sqrt{t^2+\gamma^2}$, which is given in Eq.~\eqref{e_3}.
By performing a more detailed calculation up to $\mc O(p^2)$, a better approximation is achieved, and $t$ is replaced by $\sqrt{t^2+\gamma^2}$, as given by Eqs.~\eqref{corr_diamond_3_diag} and \eqref{corr_diamond_3_off}. 
The natural correlation length in this case is $\xi=\sqrt{t^2+\gamma^2}/\gamma$.
Nonetheless, the continuum analysis accurately captures the correlations' large-distance behavior with anomalous power laws described in Eqs.~\eqref{corr_diamond_3_diag} and \eqref{corr_diamond_3_off}.

For short distances $x\ll\xi$, the correlations decay at least as fast as the conventional $1/x$ decay rate, and they exhibit qualitative agreement with the numerical results [see Fig.~\ref{fig:corr_diamond}(c)]. 
Note that \emph{short distance} is defined relative to the correlation length, and $n$ can be large, as shown in the figure, as long as $\xi\gg n$.
The short-distance behavior may be linked to the Hermitian limit, which is obtained by setting $\gamma\to0$ or $\xi\to \infty$. In the Hermitian limit, only two correlations remain non-vanishing, specifically, 
\begin{equation}
    \avg{\psi_1^\dag(x)\psi_2^{}(0)}=\avg{\psi_2^\dag(x)\psi_3^{}(0)}=-i\sqrt{2}/4\pi x.
\end{equation}

When considering $N>3$, only numerical simulations are used to investigate the power-law decay of correlations.
To generate higher EPs, a strip of the diamond lattice with increasing width is utilized, with the weights of hopping, dissipation rates, and chemical potential adjusted accordingly [as given by Eq.~\eqref{ham_diamond}].
For instance, the Bloch Hamiltonian of the model accommodating \ep{4} is expressed as
\begin{equation} 
    h_k = \pmat{
    \frac32\mu & \frac{\sqrt 3}{2}f_{k} & 0 & 0\\
    \frac{\sqrt 3}{2}f_{-k} & \frac12\mu & f_k & 0 \\
    0 & f_{-k} & -\frac12\mu & \frac{\sqrt 3}{2}f_k\\
    0 & 0 & \frac{\sqrt 3}{2}f_{-k} & -\frac32\mu 
    },
\end{equation}
with $f_k$ from Eq.~\eqref{fk}, and $\mu=\pm \gamma$.
The spectrum has four bands $E_m$~\eqref{e_m}, with spin quantum number $m\in\{\pm\frac12,\pm\frac32\}$.
In Fig.~\ref{fig:corr_diamond}(d) we show the expected $1/n^2$ and $1/n^3$ decay of correlations in the large-distance limit.

We compute numerically all distinct $N(N+1)/2$ correlations characteristic for an $N$ model with $N=2,\ldots,7$, and we fit them  with a power law which allows us to extract the decay exponents.
A representation of our findings is shown in Fig.~\ref{fig:power}(b).
The results indicate that the power-law critical behavior persists at higher-order exceptional points.
Moreover, correlations are never suppressed more than $1/n^3$.
Beyond $N=7$, calculations indicate the same conclusion, but numerical errors in fitting the large-distance behavior are enhanced.

\subsection{Charge and current densities}
\label{sec:diamond_charge}
The charge density is obtained from the diagonal correlation functions evaluated at $n=0$ or $x=0$, and it has the same profile in every unit cell in the infinite or periodic lattice due to translation symmetry.
However, with increasing gain-loss rates $\gamma$, the charge density varies within the unit cell.
This is a consequence of the model being tuned to the EP, which results in the chemical potential being pinned to the dissipation rates $|\mu|=\gamma$, causing the chemical potential to change sign from positive to negative within the unit cell, as described by Eq.~\eqref{coeffs}.
As a result, a charge imbalance between the two sides of the diamond lattice stripes is generated.

\begin{figure}[t]
\includegraphics[width=\columnwidth]{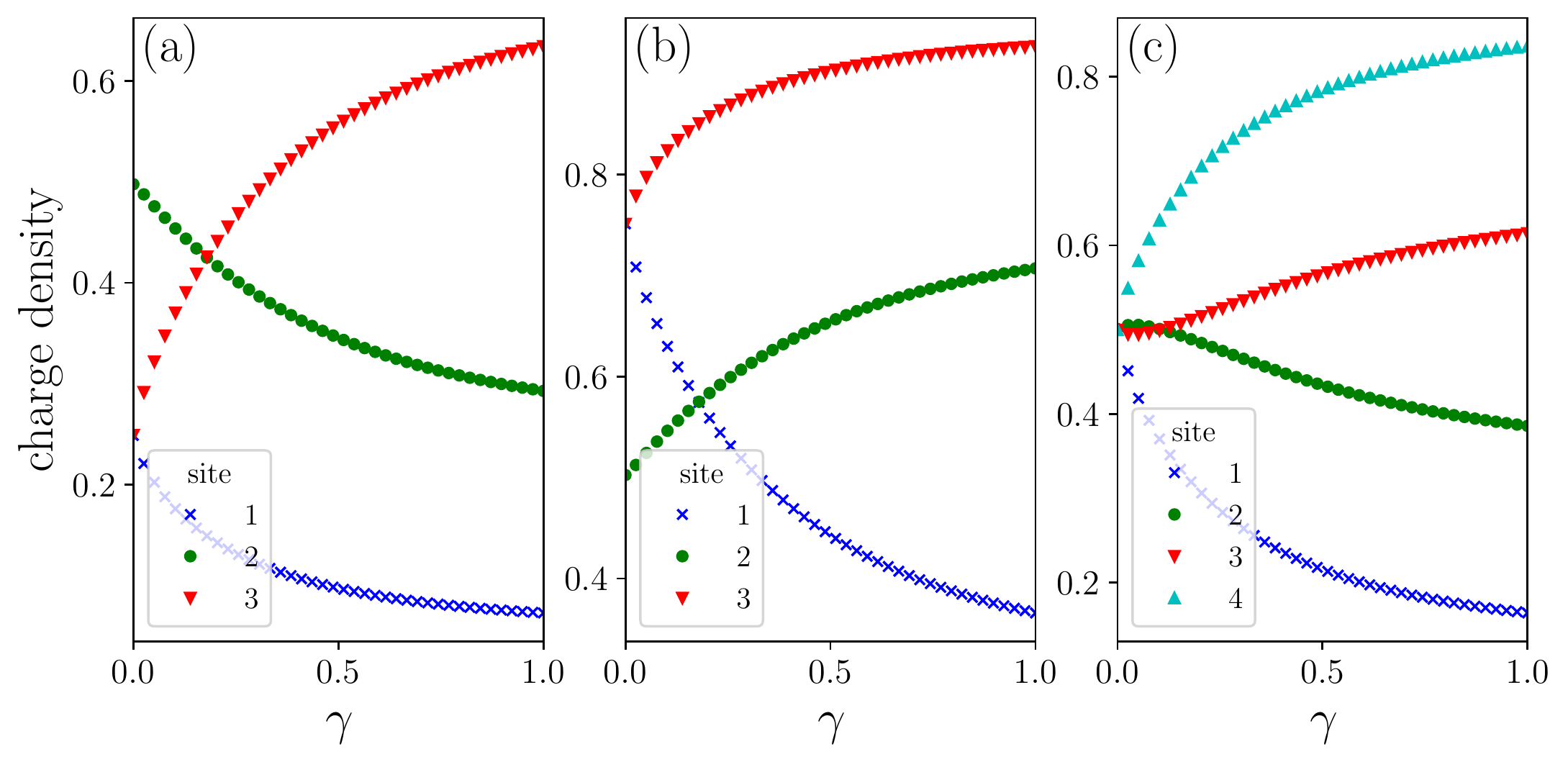}
\caption{Charge-density imbalance with increasing rates $\gamma$ in the diamond lattice with \ep{3} at (a) $\nu=1/3$ and (b) $\nu=2/3$, and (c) with \ep{4}, $\nu=1/2$.
The charge density is shown on each site in the unit cell numbered according to the convention shown in Fig.~\ref{fig:diamond_latt}.
}
\label{fig:diamond_charge_density}
\end{figure}

In continuum, the correlators in Eq.~\eqref{diamond_cont_intra_3} in the limit $x\to0$ are divergent. 
The imbalance between the density at sites in the unit cell is determined in continuum at $x=0$ by considering a cutoff $1/\delta$, which plays the role of lattice spacing.  
Then the density at $\mu=\gamma$ is approximated to leading terms in $1/\delta$,
\begin{equation}
\avg{\psi_{1,3}^\dag(0)\psi^{}_{1,3}(0)} \sim \frac{1}{4\pi\delta} \pm \frac{1}{\pi\xi}
\ln(\frac{2\delta}{\xi}).
\end{equation}
This shows that from an initial equiprobable charge distribution on the two edge sites of the lattice in the Hermitian system $\gamma\to 0$, an imbalance develops at finite $\gamma$.
The numerical evolution of charge density in a lattice unit cell as $\gamma$ increases is displayed in Fig.~\ref{fig:diamond_charge_density}(a) and (b), for $\nu=1/3$ and $\nu=2/3$, respectively, and it corroborates the analytical result at large $\xi$, small $\gamma$.

At $N>3$, charge densities at large rates $\gamma$ obtained from correlation functions continue to show charge imbalanced in the unit cell.
A numerical example is shown in Fig.~\ref{fig:diamond_charge_density}(c) for $N=4$ case. From an initial equiprobable distribution of two particles on four sites in the Hermitian $\gamma=0$ model's unit cell, a strong imbalance develops at finite $\gamma$ with a density migrating to the negative chemical potential lattice edge.

The local charge density $\rho(x,t)$ varies in time according to the non-Hermitian Heisenberg equation for expectation values~\cite{Dattoli1990,Graefe2008,Sticlet2022},
\begin{equation}\label{rho_t}
i\pd_t\avg{\rho}=\avg{H\rho-\rho H^\dag} - \avg{H-H^\dag}\avg{\rho},
\end{equation}
where the  expectation value is taken with respect to the ground state of the system. 
The right-hand side (RHS) in Eq.~\eqref{rho_t} is decomposed into a unitary and a non-unitary evolution,
\begin{equation}
i\pd_t\avg{\rho}=\avg{[\frac{H+H^\dag}{2},\rho]}+\avg{\{\frac{H-H^\dag}{2},\rho\}}
-\avg{H-H^\dag}\avg{\rho},
\end{equation}
where commutators and anticommutators are denoted with the usual symbols $[\ldots]$ and $\{\ldots\}$, respectively.
The last two terms represent the non-unitary evolution and enter the continuity equation as source-sink terms $s$,
\begin{equation}
\pd_t\avg{\rho}=-\bm\nabla \avg{j}+s.
\end{equation}
Therefore, the current in the ground state at $t=0$ is determined as
\begin{equation}\label{curr}
\bm \nabla \avg{j} = i\avg{[\frac{H+H^\dag}{2},\rho]}.
\end{equation}
Discretizing the equation on a lattice, using the tight-binding model from Eq.~\eqref{ham_diamond}, yields
\begin{equation}
\avg{j^{jj-1}_{nn+1}-j^{j+1,j}_{n-1,n}+j^{jj+1}_{nn}-j_{nn}^{j-1,j}}=i\avg{[\frac{H+H^\dag}{2},c^\dag_{j,n}c^{}_{j,n}]},
\end{equation}
where the left-hand side (LHS) represent the bond current densities. The respective bond is specified by using lower indices to denote the unit cells and upper indices, to denote the sites in the unit cell.
After solving the commutation relation on the RHS, one identifies the following currents inside the unit cell and between near-neighbor unit cells:
\begin{eqnarray}
\avg{j^{jj+1}_{nn}} &=& \frac{itS^x_{jj+1}}{2}\avg{c^\dag_{j+1,n}c^{}_{j,n}-\hc},\\
\avg{j^{jj-1}_{nn+1}} &=& \frac{itS^x_{jj-1}}{2}\avg{c^\dag_{j-1,n+1}c^{}_{j,n}-\hc},
\notag
\end{eqnarray}
respectively.

\begin{figure}
\includegraphics[width=\columnwidth]{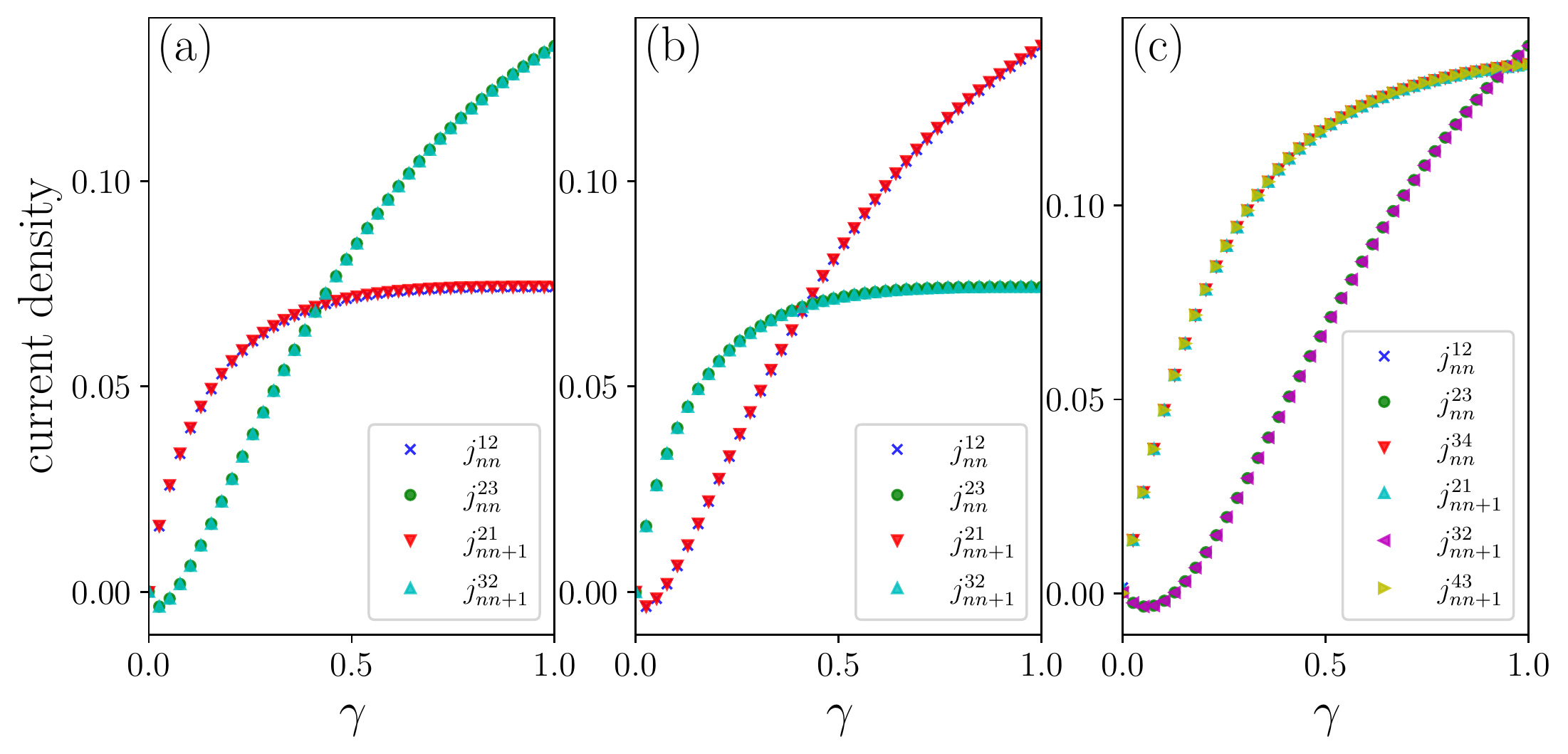}
\caption{Bond current densities in the ground state as a function of dissipation rates $\gamma$ for (a) $N=3$, $\nu=1/3$, (b) $N=3$, $\nu=2/3$, and (c) $N=4$, $\nu=1/2$. Currents $j$ and rates $\gamma$ are in units of $t$.}
\label{fig:diamond_current_density}
\end{figure}

Numerical results for $N=3$ and $N=4$ lattices are displayed in Fig.~\ref{fig:diamond_current_density} and are generic for \ep{N} diamond lattice models.
As expected, currents in the Hermitian limit $\gamma=0$ are zero. At finite $\gamma$, one may well understand the results by considering the lattice made of parallel chains of the \ep{2} models.
A current in the ground state develops in each chain, flowing to the right for almost all $\gamma$. 
Moreover, the bond currents are identical in each chain $j^{jj+1}_{nn}=j^{j+1,j}_{nn+1}$, for each chain $j$. 
For even $N$, additional reflection symmetries lead to identical currents on chains on opposite sides of the lattice, as seen for $N=4$. 
This is also expected from the symmetrical charge imbalance seen in even $N$ models that are always at half-filling [e.g.,~Fig.~\ref{fig:diamond_charge_density}(c)].

\subsection{Entanglement entropy}
\label{sec:ee_diamond}
Typically, for a Hermitian system at a quantum critical point, the bipartite entanglement entropy exhibits a logarithmic increase as the size of a subsystem grows~\cite{Eisert2010,Calabrese2004}.
This picture is different in a non-Hermitian \pt\ symmetric system supporting an \ep{2}~\cite{Dora2022} as it was shown that there is a crossover from a logarithmic growth to a constant entanglement defined by the correlation length.
This difference occurs due to the quantum Zeno effect, which arrests the propagation of correlation due to continuous monitoring from the environment in the non-Hermitian case.
This results in correlations that decay faster than in critical non-interacting Hermitian models beyond the non-Hermitian correlation length. As a result, regions separated by distances larger than the correlation length become disentangled. The only entangled degrees of freedom remain at the subsystem's borders.
Since the model is one-dimensional, the border is zero-dimensional, and the entropy is expected to become constant, $S\sim l^0$, with $l\gg \xi$, the subsystem size.
In this section, we expand the study to investigate the case of a \pt-symmetric system supporting HOEPs, and we demonstrate that similar conclusions can be drawn.

We study the entanglement entropy in the tight-binding models~\eqref{ham_diamond} with periodic boundary conditions as a function of the size of a subsystem in a model hosting \ep{N}.
The subsystem is separated out by cutting parallel to a unit cell (see Fig.~\ref{fig:diamond_latt}), and then enlarged by adding unit cells to the right.
The entanglement entropy $S$ is then computed as a function of subsystem size in unit cells $n$ based on the eigenvalues of the correlation functions~\cite{Peschel2009} obtained in the previous section. More quantitatively,
\begin{equation}
S(n) = -\sum_{m=1}^{nN}\zeta_{m}^{(n)}\ln(\zeta_{m}^{(n)}) + (1-\zeta_{m}^{(n)})\ln(1-\zeta_{m}^{(n)}),
\end{equation}
with $\zeta_{m}^{(n)}$ the $m$th eigenvalue of the correlation matrix $C_{i,j}(n)$ of size $nN\times nN$.

\begin{figure}[t]
\includegraphics[width=\columnwidth]{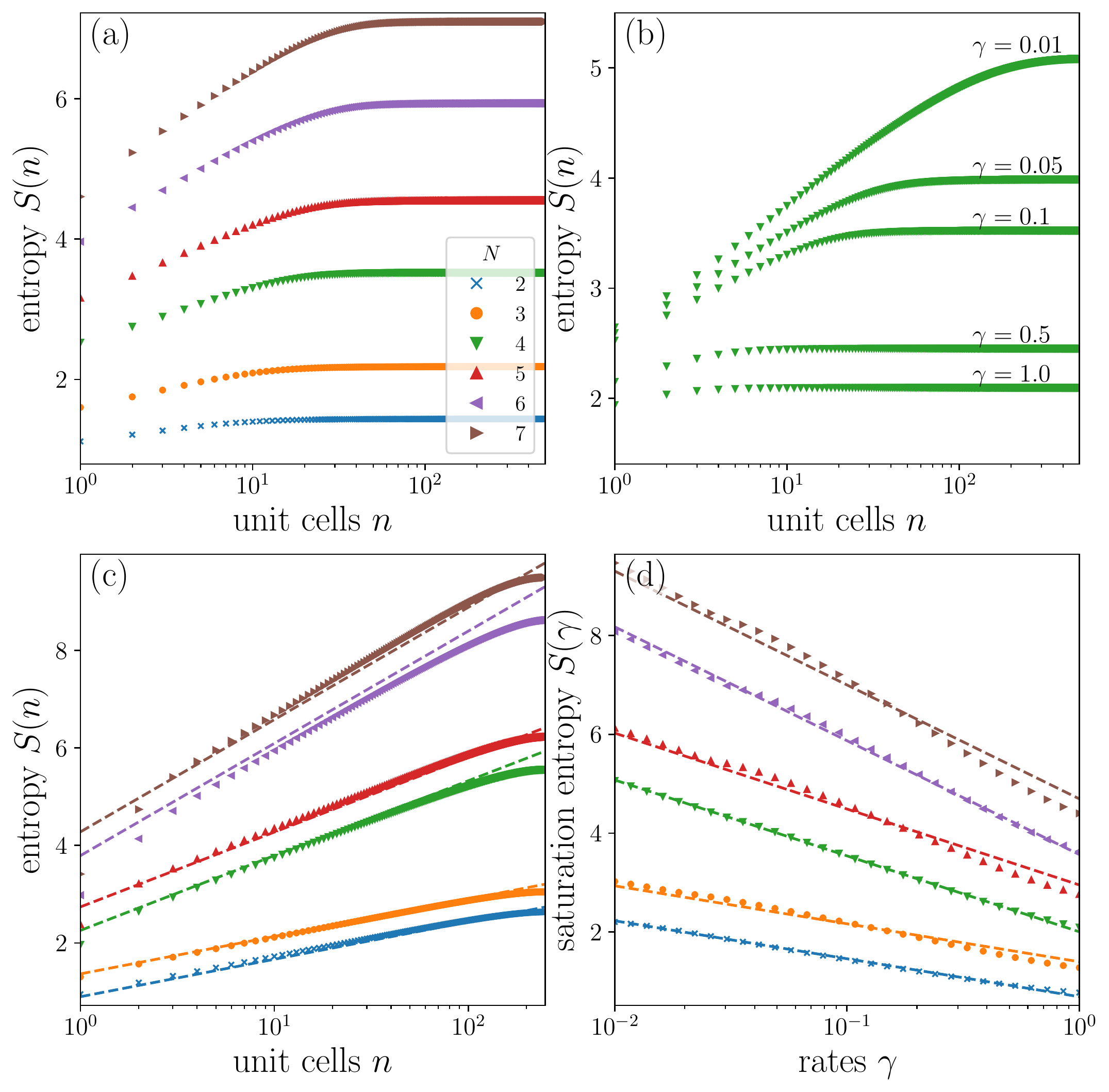}
\caption{(a) Entanglement entropy as a function of the subsystem size in unit cells $n$ at $\mu=\gamma=0.1t$. The lattice models host an \ep{N}, where $N$ ranges from  2 to 7. (b) Evolution of the entanglement entropy for $N=4$ and for different rates $\gamma$.
(c) Entanglement entropy for $n \ll t/\gamma$ (symbols) is fitted with Eq.~\eqref{s_diamond_small} (dashed lines) at $\mu=\gamma=10^{-3}t$.
(d) Saturation entropy (symbols) as a function of dissipation rates $\gamma$ for $n\gg t/\gamma$ is fitted with Eq.~\eqref{s_diamond_large} (dashed lines).
The legend is shared for all panels, and $\gamma$ is in units of $t$.}
\label{fig:ent_diamond}
\end{figure}

For models with \ep{N}, where $N$ ranges from 2 to 7, the entanglement entropy is computed numerically, and the outcomes are illustrated in Fig.~\ref{fig:ent_diamond}. 
The overall trend in all the cases is the logarithmic increase in entropy with the system's size, up to a scale of $\xi\sim t/\gamma$, beyond which the entropy reaches saturation. The crossover is observed to rise with the value of $N$, indicating that a better estimation for the correlation length is $\xi\sim Nt/\gamma$.

As shown in the previous sections, at short distances, $n\ll \xi$, the system behaves almost as a noninteracting Hermitian fermionic model~\cite{Calabrese2004} with $S(l)\sim \frac{c}{3}\log(l)$, 
with $l$ the subsystem size and with $c$ the central charge.
The entropy is well approximated by
\begin{equation}\label{s_diamond_small}
S(n\ll\xi) \approx \frac{\lfloor N/2\rfloor}{3}\log(nN)+\frac{N}{3},
\end{equation}
as it is shown in Fig.~\ref{fig:ent_diamond}(c), and $\lfloor x\rfloor$ yields the greatest integer less than or equal to $x$.
The Eq.~\eqref{s_diamond_small} indicates that the central charge $c=N/2$ for even $N$, while for odd $N$, with a flat band, $c=(N-1)/2$.
The effective size of the subsystem is given by the number of cells $n$ times the number of sites per unit cell $N$. 

On the other hand, when the subsystem size becomes large, $n\gg \xi$, the entropy saturates and displays a characteristic behavior similar to an insulator~\cite{Calabrese2004},
\begin{equation}\label{s_diamond_large}
S(n\gg\xi) \approx \frac{\lfloor N/2\rfloor}{3}\ln(\xi) + \text{const..}
\end{equation}
A fit with a single free parameter, the additive constant, is shown in Fig.~\ref{fig:ent_diamond}(d). The odd $N$ models deviate slightly from this law. For $\xi^{-1}>1$, the saturation entropy reaches its minimum value and becomes constant.

\section{Unidirectional model hosting \texorpdfstring{\ep{N}}{EPN}}
\label{sec:unidirectional}

\begin{figure}[t]
\includegraphics[width=\columnwidth]{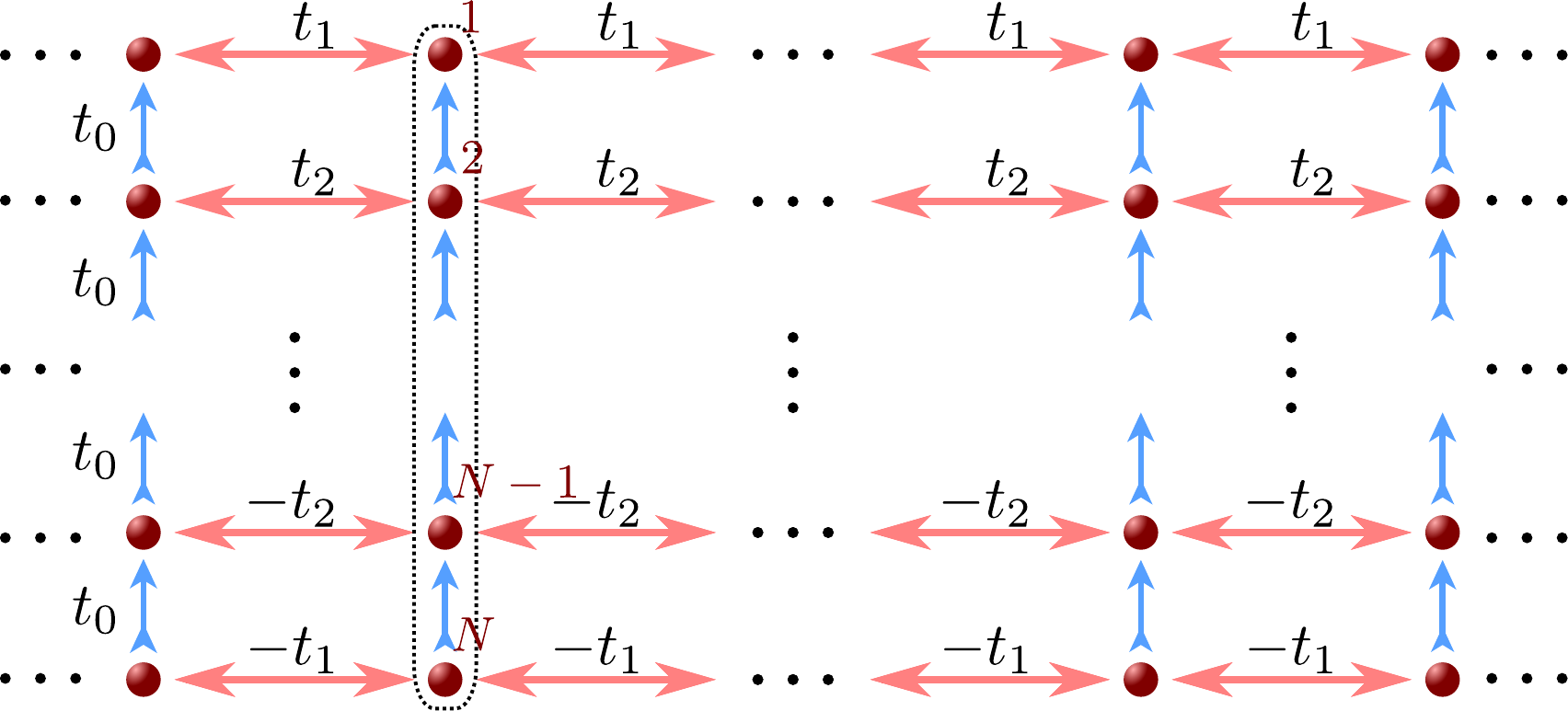}
\caption{
Tight-binding lattice model~\eqref{ham_square_tb} hosting exceptional points \ep{N}.
Reciprocal hoppings are drawn as bidirectional red arrows, while blue arrows denote unidirectional couplings of uniform amplitude $t_0$.
The sites encircled by a dotted line form a unit cell. The sites in the unit cell are counted from the top.
}
\label{fig:tb_square}
\end{figure}

In this section, we introduce a new model, characterized by the continuum Hamiltonian  near the \ep{N},
\begin{equation}
H = \sum_{i,j=1}^N\int dx\psi_i^\dag(x)h_{ij}(x)\psi_j^{}(x),
\end{equation}
where $h$ has the structure 
\begin{equation}\label{continuum_h}
h_{ij}(p) = v_i p\delta_{i,j} + \Delta\delta_{i+1,j},
\end{equation}
in momentum space. 
Indices $i,j$ take values from $1$ to $N$, and $\delta_{i,j}$, is the Kronecker symbol.
The energy spectrum of the model consists of $N$ bands, featuring an \ep{N} at momentum $p=0$, as illustrated in Fig.~\ref{fig:power}(a). While the energies are real, the model is non-Hermitian due to the presence of the off-diagonal term $\Delta$, which is reflected only in the eigenstates and not in the eigenenergies. The particle-hole symmetry of the model implies that only $\lfloor N/2\rfloor$ velocities are independent, with $v_i=-v_{N+1-i}$. Moreover, for odd $N$, a flat band with zero velocity $v_{(N+1)/2}=0$ must exist, as required by the symmetry. We choose, without loss of generality, the linearly independent velocities $v_i$ with $i<\lfloor N/2\rfloor$ to be positive and different from each other to avoid line degeneracies. This makes the spectrum qualitatively similar to the dispersion of the previous model in Eq.~\eqref{e_m}.

While our primary emphasis is on the continuum Hamiltonian, it is worth noting that the model has a lattice implementation. The proposed lattice Hamiltonian is advantageous for numerical computations, providing a means to verify the analytical findings in the continuum model, despite being more challenging to achieve experimentally. 
The tight-binding Hamiltonian presented below defines a lattice non-Hermitian model with real energy bands,
\begin{equation}\label{ham_square_tb}
H = \sum_{n=-L}^{L} \bigg[\sum_{i=1}^N
t_i(c^\dag_{i,n}c_{i,n+1}^{}+\text{H.c.}) 
+\sum_{i=1}^{N-1}t_0c^\dag_{i,n}c_{i+1,n}^{}\bigg].
\end{equation}
Creation (annihilation) operators $c_{i,n}^\dag$ ($c_{i,n}$) are labeled by the index $i$, denoting the site inside the unit cell that contains $N$ sites, and by the index $n$, designating the unit cell.
Unidirectional hopping $t_0$ occurs inside the unit cell, while reciprocal hopping $t_i$, $i>0$, occurs between sites belonging to different unit cells. 
A lattice representation of the model is displayed in Fig.~\ref{fig:tb_square}.
Such a model may be regarded either as stacking unit cells of ``one-way'' Hatano-Nelson models~\cite{Hatano1996,Feinberg1999} of finite size $N$, or as a non-Hermitian coupling through $t_0$ of $N$ Hermitian tight-binding hopping chains.
The model~\eqref{ham_square_tb} is also \pt\ symmetric, but in contrast to the previous Hamiltonian~\eqref{ham_diamond}, the model is also time-reversal $\mc T=\mc K$ symmetric. 

The Hamiltonian~\eqref{ham_square_tb} with periodic boundary conditions is diagonalized in momentum space, and it has an energy spectrum composed of $N$ bands with energies,
\begin{equation}
E_i = 2t_i\cos(k) \textmd{ for } 1\leq i \leq N,   
\end{equation}
where the lattice constant between neighboring sites is set to 1, and the spectrum is independent of $t_0$.
As in the continuum case, the particle-hole symmetry dictates that $t_i = -t_{N+1-i}$, which means that for odd values of $N$, the central hopping term $t_{(N+1)/2}$ becomes zero, resulting in the formation of a flat band in the model. If each hopping term $t_i$ is distinct, the model's spectrum will feature two $N$-fold degeneracies at momenta $k=\pm\pi/2$. 
The low-energy behavior near either of these degeneracies is characterized by the continuum Hamiltonian given in Eq.~\eqref{continuum_h}. For example, at $k=\pi/2$, the parameters of the continuum and lattice models are related as $v_i=2t_i$ and $\Delta=t_0$.

\subsection{Correlations in \texorpdfstring{\ep{2}}{EP2} model}
\begin{figure}[t]
\includegraphics[width=\columnwidth]{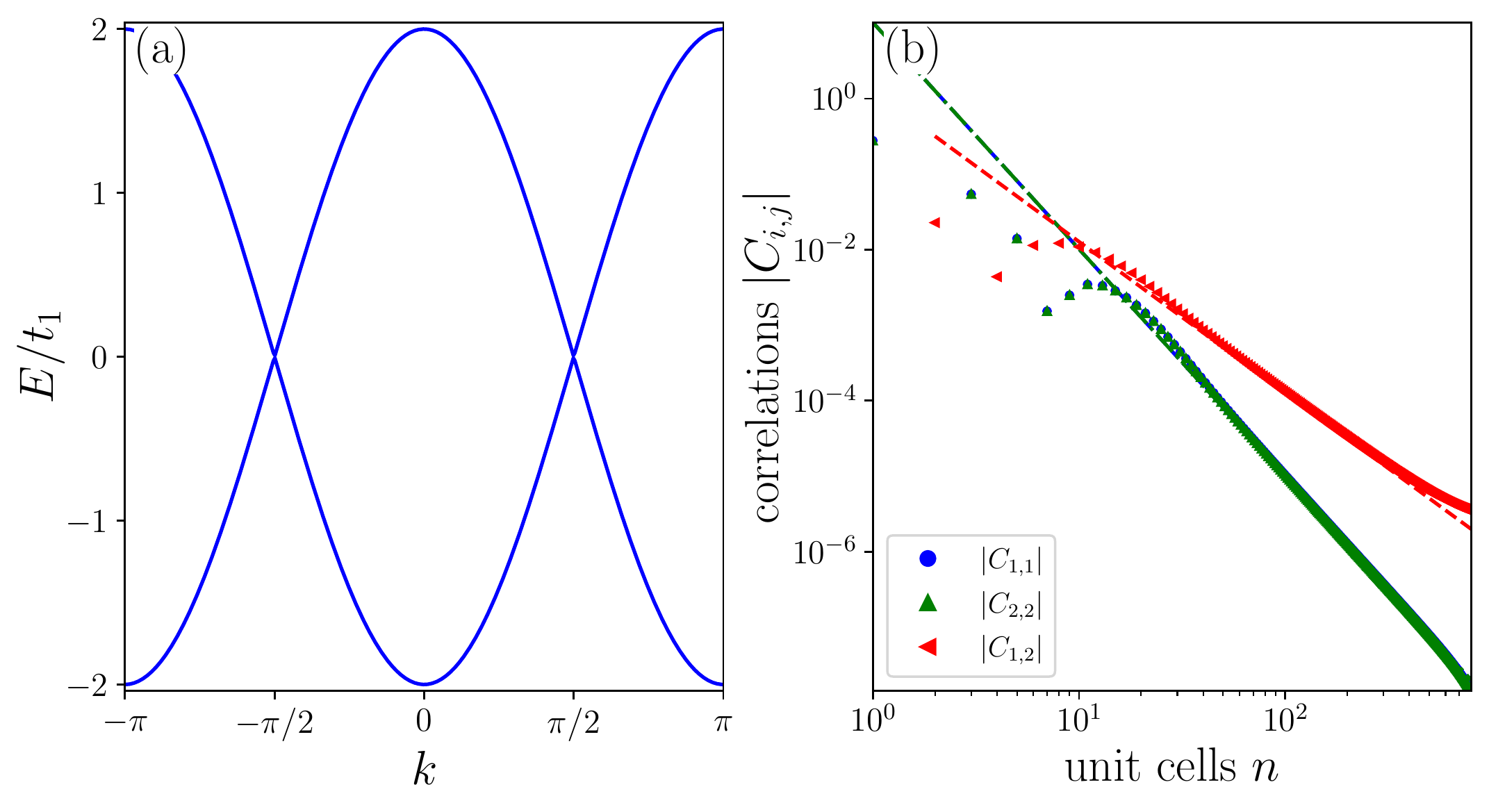}
\caption{(a) Bands of the tight-binding model realizing two \ep{2}. 
(b) The spatial variation, in unit cells $n$, for the 3 correlation functions $C_{i,j}(n)$ in the $N=2$ model~\eqref{ham_square_tb}. 
The symbols represent the numerical results, while the dashed lines denote the analytical results for  the corresponding $C_{i,j}(n)$ in Eqs.~\eqref{N2_latt}.
}
\label{fig:2_corr}
\end{figure}

The continuum Hamiltonian~\eqref{continuum_h} for $N=2$, with $v\equiv v_1$, is precisely the one reviewed in the previous section [see Eq.~\eqref{h_cont_2}], with correlations~\eqref{N2_cont}.
Additionally, we compute here the correlation functions on the lattice 
for the Hamiltonian~\eqref{ham_square_tb} in the limit of large $n$ unit cells,
\begin{eqnarray}\label{N2_latt}
\avg{c^\dag_{1,n}c_{1,0}^{}} &\sim& +\frac{32 t_1^2}{\pi t_0^2}\frac{\sin(n\pi/2)}{n^3},\notag\\
\avg{c^\dag_{2,n}c_{2,0}^{}} &\sim& -\frac{32 t_1^2}{\pi t_0^2}\frac{\sin(n\pi/2)}{n^3},\\
\avg{c^\dag_{1,n}c_{2,0}^{}} &\sim& -\frac{4t_1}{\pi t_0}\frac{\cos(n\pi/2)}{n^2}.\notag
\end{eqnarray}
The power-law decay of correlations corroborates the continuum results in Eq.~\eqref{N2_cont}.
The above results concur with the findings in Ref.~\cite{Dora2022}, where the one-dimensional tight-binding model had a single \ep{2} in the Brillouin zone (BZ). A  comparison for the correlation functions computed  numerically in a lattice model~\eqref{ham_square_tb} 
and the analytical results~\eqref{N2_latt} in the large-distance limit are presented Fig.~\ref{fig:2_corr}, showing the expected $1/n^2$ and $1/n^3$ behavior.

\subsection{Correlations in \texorpdfstring{\ep{N}}{EPN} model}
\label{sec:ep3}
The first HOEP model that corresponds to $N=3$, features a zero-energy flat band, and is analyzed analytically.
Similarly to the model in Sec.~\ref{sec:diamond_correlations}, the non-Hermitian system exhibits critical behavior that manifests as a power-law decay of correlations, but with anomalous power-law exponents. 
However, the inclusion of a third energy level results in even stronger suppression of correlations compared to the case in which there are only two energy levels.
\begin{figure}[t]
    \includegraphics[width=\columnwidth]{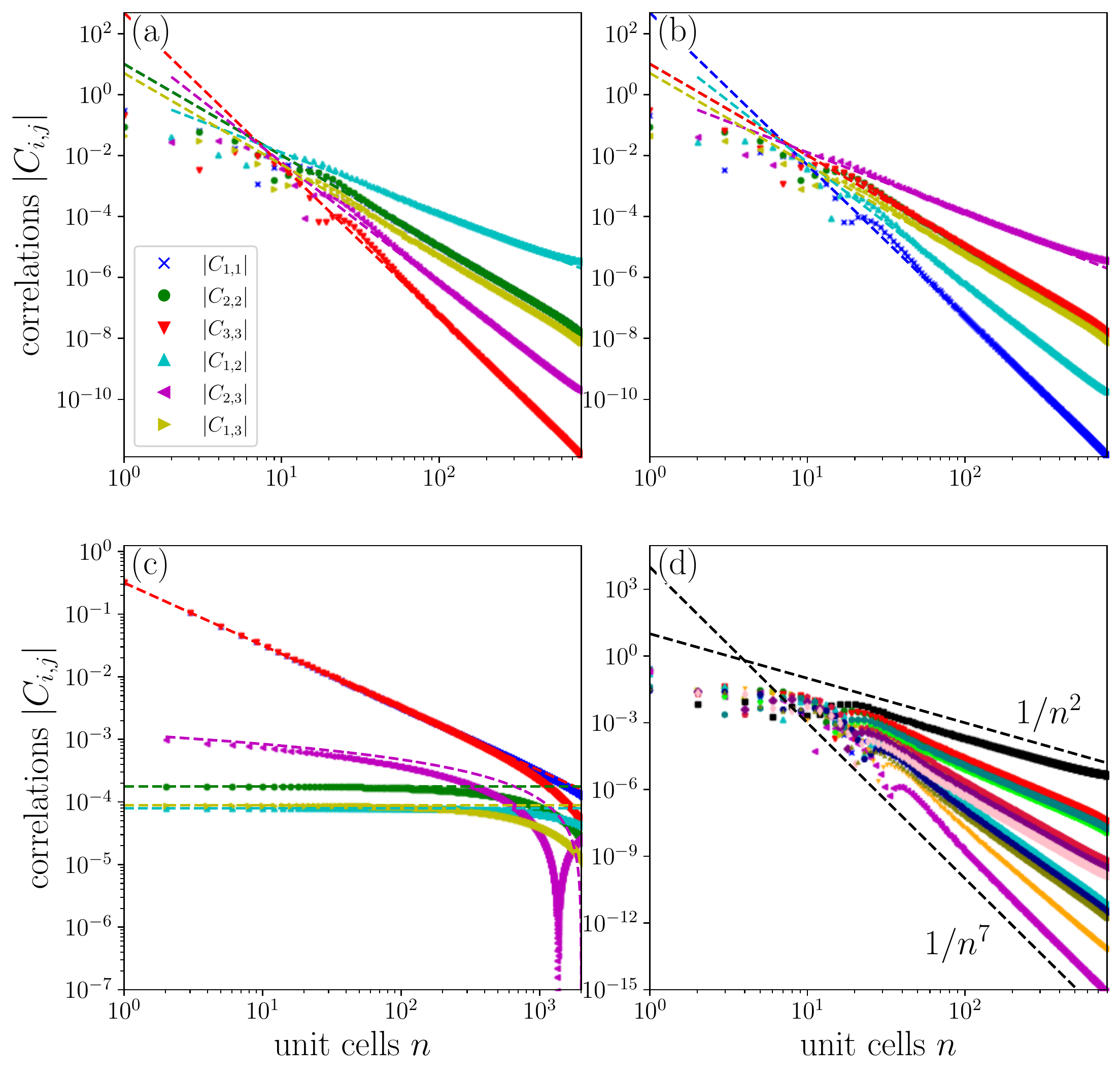}
    \caption{The spatial variation, in unit cells $n$, of the 6 correlation functions $C_{i,j}(n)$ in the $N=3$ model~\eqref{ham_square_tb} at filling (a) $\nu=1/3$ and (b) $\nu=2/3$ for $t_0=t$. 
    Symbols represent the numerical results, while the dashed lines are the analytical results for the corresponding $C_{i,j}(n)$ in Eqs.~\eqref{intra13}, \eqref{inter13}, and \eqref{phs_latt_3}.
    Correlations at $\nu=1/3$ for $t_0=10^{-3}t$, in the short-distance limit, are fitted with the corresponding continuum model results Eqs.~\eqref{ham2_cont_intra13} and~\eqref{ham2_cont_inter13}.
    The legend is shared among panels (a), (b), and (c).
    The matching color of a symbol from numerics and the dashed line from analytics identify the same correlation function.
    (d) All 15 distinct correlation functions for $N=5$ exhibit stronger suppression in the large-distance limit, from $1/n^2$ up to $1/n^7$, as indicated by the dashed line.
    }
    \label{fig:3_corr}
\end{figure}

The continuum model~\eqref{continuum_h} for \ep{3} reads
\begin{equation}\label{ham2_cont_3}
h = \pmat{vp& \Delta & 0\\ 0 & 0 & \Delta \\ 0 & 0  &- vp}.
\end{equation}
Diagonalizing the matrix Hamiltonian, the energies are $E_1=vp$, $E_2=0$, and $E_3=-vp$, respectively.
In an equivalent Hermitian model with $\Delta=0$, the three energy bands are uncoupled. 
The only existing correlations are between the right-moving fields $\avg{\psi_1^\dag(x) \psi_1^{}(0)}$ and the left-moving fields $\avg{\psi_3^\dag(x) \psi_3^{}(0)}$. Moreover, there are no correlations in the flat band, as $\avg{\psi_2^\dag(x) \psi_2^{}(0)}$ equals zero. As a result, the correlation functions are either zero or exhibit the conventional $1/x$ decay in one dimension.

In the non-Hermitian model, the non-Hermitian mass $\Delta$ couples all the bands, leading to non-trivial correlations among all fields $\psi_i$. In the following, we calculate correlation functions for both $1/3$ or $2/3$ filling, corresponding to an empty or filled flat band, respectively.

At $1/3$ filling, the diagonal correlation functions read
\begin{eqnarray}\label{ham2_cont_intra13}
\avg{\psi_1^\dag(x)\psi^{}_1(0)} &=& 
\frac{i\Delta}{2\pi v}\times
\begin{cases}
\frac{v}{\Delta x} & x\ll \frac{v}{\Delta},\\
-(\frac{2v}{\Delta x})^3& x\gg \frac{v}{\Delta},
\end{cases}\notag\\
\avg{\psi_2^\dag(x)\psi^{}_2(0)} &=& \frac{i\Delta}{2\pi v}\times
\begin{cases}
\frac{\sqrt 2\pi}{4i}& x\ll \frac{v}{\Delta},\\
(\frac{2v}{\Delta x})^3& x\gg \frac{v}{\Delta},
\end{cases}
\\ 
\avg{\psi_3^\dag(x)\psi^{}_3(0)} &=& 
-\frac{i\Delta}{2\pi v}\times
\begin{cases}
\frac{v}{\Delta x} & x\ll \frac{v}{\Delta},\\
3 (\frac{2v}{\Delta x})^5& x\gg \frac{v}{\Delta},
\end{cases}
\notag
\end{eqnarray}
and the off-diagonal ones,
\begin{eqnarray}\label{ham2_cont_inter13}
\avg{\psi_1^\dag(x)\psi^{}_2(0)}
&=&\frac{\Delta}{4\pi v}\times
\begin{cases}
    -1 & x\ll \frac v\Delta, \\
    (\frac{2v}{\Delta x})^2 & x\gg \frac v \Delta, 
\end{cases}
\notag\\
\avg{\psi_1^\dag(x)\psi^{}_3(0)}
&=&\frac{\Delta}{4\pi v}\times
\begin{cases}
    \frac{\sqrt 2\pi}{4} & x\ll \frac v\Delta,\\
    i(\frac{2v}{\Delta x})^3 & x\gg \frac v\Delta,
\end{cases}\\
\avg{\psi_2^\dag(x)\psi^{}_3(0)} 
&=&
\frac{\Delta}{4\pi v}\times
\begin{cases}
    2\ln(\frac{\Delta x}{v}) & x\ll \frac v\Delta,\\
    -3 (\frac{2v}{\Delta x})^4 & x\gg \frac v\Delta.
\end{cases}\notag
\end{eqnarray}
Instead of the conventional $1/x$ decay, the correlations are further suppressed in the non-Hermitian model.
At short distances, $\avg{\psi_2^\dag(x)\psi_2^{}(0)}$ and the off-diagonal correlators $\avg{\psi_i^\dag(x)\psi_{j\neq i}^{}(0)}$ vanish similar to the Hermitian limit, as $\Delta\to 0$. 

At $\nu = 2/3$ filling, the diagonal correlations are
\begin{eqnarray}\label{intra23}
    \avg{\psi_1^\dag(x)\psi^{}_1(0)} &=& 
    \frac{i\Delta}{2\pi v}\times
    \begin{cases}
        \frac{v}{\Delta x} & x\ll \frac{v}{\Delta},\\
        3 (\frac{2v}{\Delta x})^5& x\gg \frac{v}{\Delta},
    \end{cases}
    \notag\\
    \avg{\psi_2^\dag(x)\psi^{}_2(0)} &=& -\frac{i\Delta}{2\pi v}\times
    \begin{cases}
    \frac{\sqrt 2\pi}{4i}& x\ll \frac{v}{\Delta},\\
    (\frac{2v}{\Delta x})^3& x\gg \frac{v}{\Delta},
    \end{cases}
    \\ 
    \avg{\psi_3^\dag(x)\psi^{}_3(0)} &=& 
    \frac{i\Delta}{2\pi v}\times
    \begin{cases}
    -\frac{v}{\Delta x} & x\ll \frac{v}{\Delta},\\
    (\frac{2v}{\Delta x})^3& x\gg \frac{v}{\Delta},
    \end{cases}\notag
\end{eqnarray}
while the off-diagonal ones read,
\begin{eqnarray}\label{inter23}
    \avg{\psi_1^\dag(x)\psi^{}_2(0)}
    &=&
    \frac{\Delta}{4\pi v}\times
    \begin{cases}
        2\ln(\frac{\Delta x}{v}) & x\ll \frac v\Delta,\\
        -3 (\frac{2v}{\Delta x})^4 & x\gg \frac v\Delta,
    \end{cases}\notag\\
    \avg{\psi_1^\dag(x)\psi^{}_3(0)}
    &=&\frac{\Delta}{4\pi v}\times
    \begin{cases}
        -\frac{\sqrt 2\pi}{4} & x\ll \frac v\Delta,\\
        -i(\frac{2v}{\Delta x})^3 & x\gg \frac v\Delta,
    \end{cases}\\
    \avg{\psi_2^\dag(x)\psi^{}_3(0)}
    &=&
    \frac{\Delta}{4\pi v}\times
    \begin{cases}
        -1 & x\ll \frac v\Delta, \\
        (\frac{2v}{\Delta x})^2 & x\gg \frac v \Delta.
    \end{cases}
    \notag
\end{eqnarray}
When the filling is $\nu=2/3$, two bands in the Fermi sea are filled up the zero energy corresponding to \ep{N}, namely, a dispersing mode and the flat band. Since the Hamiltonian is non-Hermitian, the corresponding right eigenvectors are not orthogonal. 
Consequently, an orthogonalization procedure is required on the eigenvectors in the Fermi sea to compute expectation values correctly  and then take the trace over the occupied states~\cite{Herviou2019,Sticlet2022}. 
Alternatively, it is more convenient to use the particle-hole symmetry to directly determine the correlations from the known $\nu=1/3$ case
(see App.~\ref{sec:phs_flat}),
\begin{equation}\label{phs_cont_main}
C^{\nu=2/3}_{i,j}(x) = (-1)^{i+j+1}C^{\nu=1/3}_{4-j,4-i}(x).
\end{equation}

The lattice model is analytically solved in the limit of large distances, making it possible to compare directly the decay of the correlations with those of the continuum model.
We find diagonal correlators at $\nu=1/3$ filling over a distance of $n$ unit cells,
\begin{eqnarray}\label{intra13}
\avg{c_{1,n}^\dag c^{}_{1,0}} &\sim& -\frac{32 t^2_1}{\pi t_0^2}\frac{\sin(n\pi/2)}{n^3},\notag\\
\avg{c_{2,n}^\dag c^{}_{2,0}} &\sim& +\frac{32 t^2_1}{\pi t_0^2}\frac{\sin(n\pi/2)}{n^3},\\
\avg{c_{3,n}^\dag c^{}_{3,0}} &\sim& -\frac{1536 t^4_1}{\pi t_0^4}\frac{\sin(n\pi/2)}{n^5}\notag.
\end{eqnarray}
The off-diagonal correlators read
\begin{eqnarray}\label{inter13}
\avg{c_{1,n}^\dag c^{}_{2,0}} &\sim& -\frac{4t_1}{\pi t_0}\frac{\cos(n\pi/2)}{n^2},\notag\\
\avg{c_{1,n}^\dag c^{}_{3,0}} &\sim& +\frac{16t^2_1}{\pi t_0^2}\frac{\sin(n\pi/2)}{n^3}, \\
\avg{c_{2,n}^\dag c^{}_{3,0}} &\sim& +\frac{192 t^3_1}{\pi t_0^3}\frac{\cos(n\pi/2)}{n^4}.
\notag
\end{eqnarray}
The correlations at $\nu=2/3$ are obtained either by direct evaluation or by using the same particle-hole symmetry present in the model. In Appendix~\ref{sec:example}, we explicitly demonstrate that the two approaches converge in obtaining a correlator.

As in the previous section~\ref{sec:spin}, the analytical results clearly demonstrate the emergence of a correlation length $\xi\simeq v/\Delta$, which indicates a transition from an almost Hermitian behavior when $\Delta\to 0$ to a large-distance regime $x\gg \xi$ where the correlations decay according to a power-law. In contrast to the models discussed in Sec.~\ref{sec:spin}, the anomalous decay exponent in the power laws is proportional to the EP order. For instance, the most suppressed correlator at $\nu=1/3$ follows $C_{3,3}(x)\sim 1/x^5$. This correlation decays faster than the ones observed in the previous section.

Figures~\ref{fig:3_corr}(a),~\ref{fig:3_corr}(b), and~\ref{fig:3_corr}(c) compare the numerical calculation of correlations with the analytical results described above. The large-distance behavior $x\gg \Delta/v$ (or $n\gg t_0/2t_1$ on the lattice) is evident in Figs.~\ref{fig:3_corr}(a) and (b) at $\nu=1/3$ and $\nu=2/3$, respectively. The analytical lattice results~\eqref{intra13} and \eqref{inter13} accurately predict the power-law decay of correlations with $1/x^a$, where $a$ is in the set $\{2,3,4,5\}$. 
The panels also illustrate the expected interchange in the behavior of correlation functions under the particle-hole symmetry that relates cases with filling $\nu=1/3$ to $\nu=2/3$.
The short-distance $x\ll \Delta/v$ behavior of correlations is shown in Fig.~\ref{fig:3_corr}(c).
Here, the comparison is made with the analytical results in the continuum model at $\nu=1/3$ [Eqs.~\eqref{ham2_cont_intra13} and~\eqref{ham2_cont_inter13}]. The latter results describe a single \ep{3}, and, since the lattice hosts two \ep{N}, the correlations in the lattice are double the results in the continuum limit.

Numerical computations of correlations for the lattice model~\eqref{ham_square_tb} are performed for $N>3$. 
The $N$ hoppings $t_i$ are uniformly distributed in the interval $[-t,t]$, and the computations are done in units of $t$. 
In numerics, $t_0$ is on the order of $t$, making the asymptotic large-distance $n\gg t/t_0$ regime quickly attainable after a few sites. An instance of $N=5$ at $\nu=2/5$ is displayed in Fig.~\ref{fig:3_corr}(d), where correlations are seen to be suppressed with different power laws $1/n^{a}$, with $a$ an integer in between 2 and 7. For general $N$, the correlations' evolution is fitted with the power-law decay $1/n^{a}$, and the decay exponents are extracted and displayed in Fig.~\ref{fig:power}(c). The results reveal an increasing suppression of correlations with the order of the EP. In all cases, the most suppressed correlator when the flat band is empty, $\nu=(N-1)/2N$, is the one corresponding to sites on the chain $N$ that is most depleted by particles.
The power-law exponent varies linearly with $N$ as
\begin{equation}
\avg{\psi_N^\dag(x)\psi_N^{}(0)}_{\nu=\frac{N-1}{2N}} \sim 
\begin{cases}
x^{-N-2}& N \text{ odd},\\
x^{-N-1}& N \text{ even}.
\end{cases}
\end{equation}
For a filled flat band, the most suppressed correlators always include $\avg{\psi_1^\dag(x)\psi_1^{}(0)}$.

\subsection{Charge and current densities}
We examine the charge density of the system using the correlation functions, as in Sec.~\ref{sec:spin}.
Because of translational symmetry, the density profile is the same in all unit cells.
Similar to the diamond lattice previously discussed, charge imbalance arises in the non-Hermitian system.
However, unlike in the diamond lattice, this is not due to a chemical potential difference between the lattice's sides, but rather to the unidirectional hopping $t_0$.

\begin{figure}[t]
\includegraphics[width=\columnwidth]{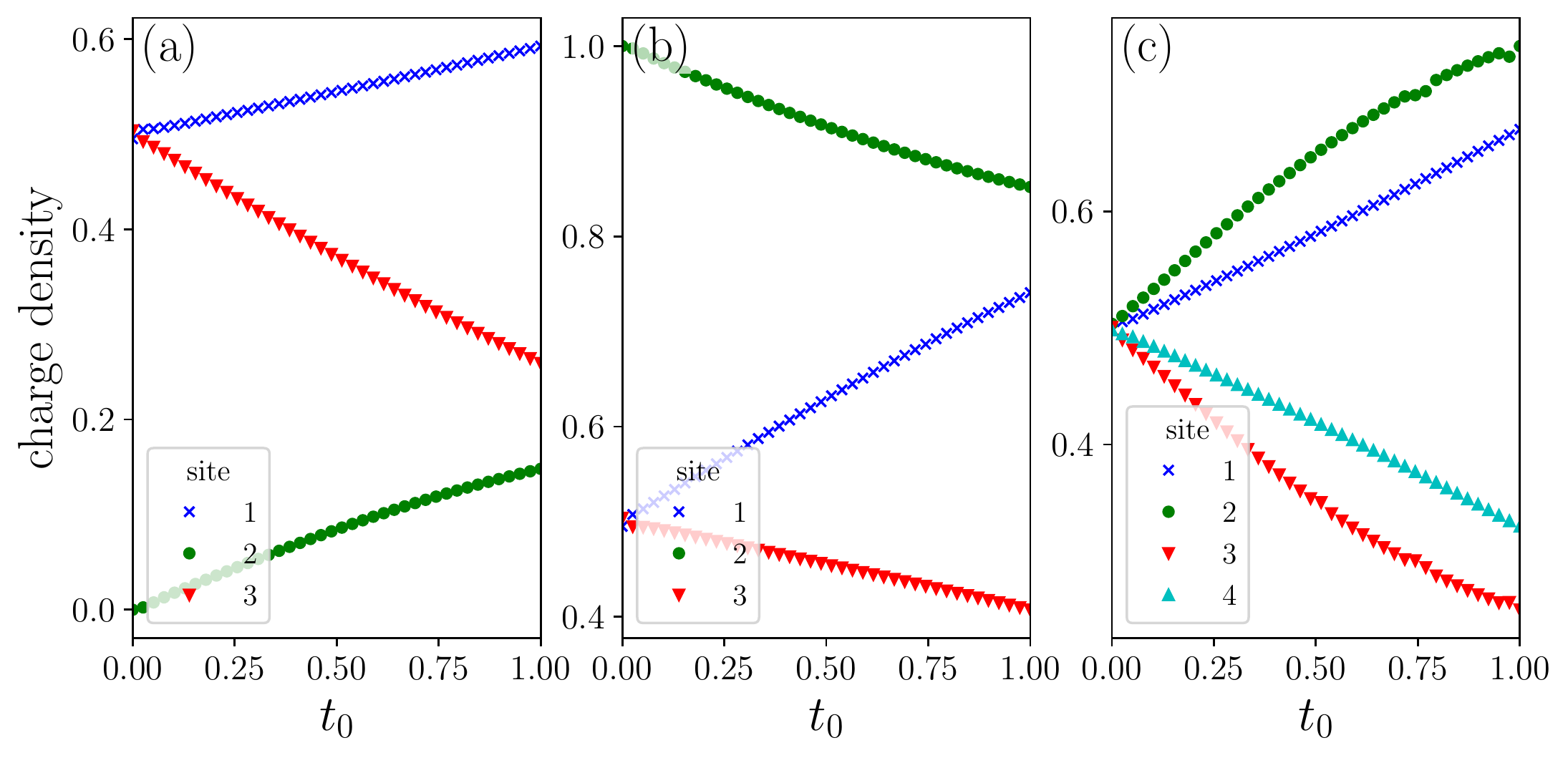}
\caption{The dependence of the charge density on the amplitude $t_0$ of the unidirectional hopping.
Here, $t_0$ is measured in units of the maximal reciprocal hopping amplitude $t$.
Panels (a) and (b) present the case $N=3$ at filling $\nu=1/3$ and $\nu=2/3$, respectively, while (c) shows the case $N=4$, $\nu=1/2$.
}
\label{fig:square_charge_density}
\end{figure}

In Fig.~\ref{fig:square_charge_density}(a) and (b), we present the results for the $N=3$ lattice.
To obtain the charge imbalance analytically, we take the limit $x\to 0$ in the diagonal correlation functions, with a cutoff $1/\delta$ curing the divergent correlations.
For $\nu=1/3$, the charge density on the edges of the lattice, sites one and three in the unit cell, is determined by the two leading terms in $1/\delta$,
\begin{eqnarray}
\avg{\psi^\dag_1(0)\psi_1(0)} \sim\frac{1}{2\pi\delta}+\frac{\sqrt2\Delta}{16v},\nonumber\\
\avg{\psi^\dag_3(0)\psi_3(0)} \sim\frac{1}{2\pi\delta}-\frac{3\sqrt2\Delta}{16v}.
\end{eqnarray}
Together with $\avg{\psi^\dag_2(0)\psi_2(0)}=\sqrt2\Delta/8v$ from Eq.~\eqref{ham2_cont_intra13}, it is apparent that analytics correctly describe the linear dependence of the density on the unidirectional hopping, at weak $\Delta$ or $t_0$, from Fig.~\ref{fig:square_charge_density}(a).
The phenomenon of charge-density imbalance due to unidirectional hopping $t_0$ is observed in the charge density of the lattice at $N>3$ as well. This is in line with the behavior expected from the unidirectional Hatano-Nelson model that characterizes the unit cell. However, the presence of transversal hopping $t_i$, with larger values at the edges of the lattice, can lead to the charge density having a maximum away from the last site of the lattice, as seen in Fig.~\ref{fig:square_charge_density}(b) for the $N=4$ lattice. Nonetheless, for sufficiently large unidirectional hopping $t_0\gg t$, the transversal skin effect becomes more prominent, and the charge density is maximal at the edge sites.

The bond charge current densities are determined according to the theory in Sec.~\ref{sec:diamond_charge},
\begin{eqnarray}\label{sq_currents}
\avg{j^{jj+1}_{nn}} &=& \frac{it_0}{2}\avg{c^\dag_{j+1,n}c^{}_{j,n} - \hc},\\
\avg{j^{jj}_{nn+1}} &=& it_j\avg{c^\dag_{j,n+1}c^{}_{j,n} - \hc},\notag
\end{eqnarray}
with $\avg{j^{jj+1}_{nn}}$, the transverse current inside a given unit cell $n$, and $\avg{j^{jj}_{nn+1}}$, the longitudinal current between sites $j$ in near-neighbor unit cells.
It is noteworthy that, as also seen in analytics, the lattice correlation matrix is real and symmetric (Hermitian), leading to cancellation of the two terms in each Eq.~\eqref{sq_currents}. 
Therefore, in contrast to the previous models in Sec.~\ref{sec:spin}, there are no currents in the ground state.

\subsection{Entanglement entropy}
\label{sec:ee_square}
\begin{figure}[t]
\includegraphics[width = \columnwidth]{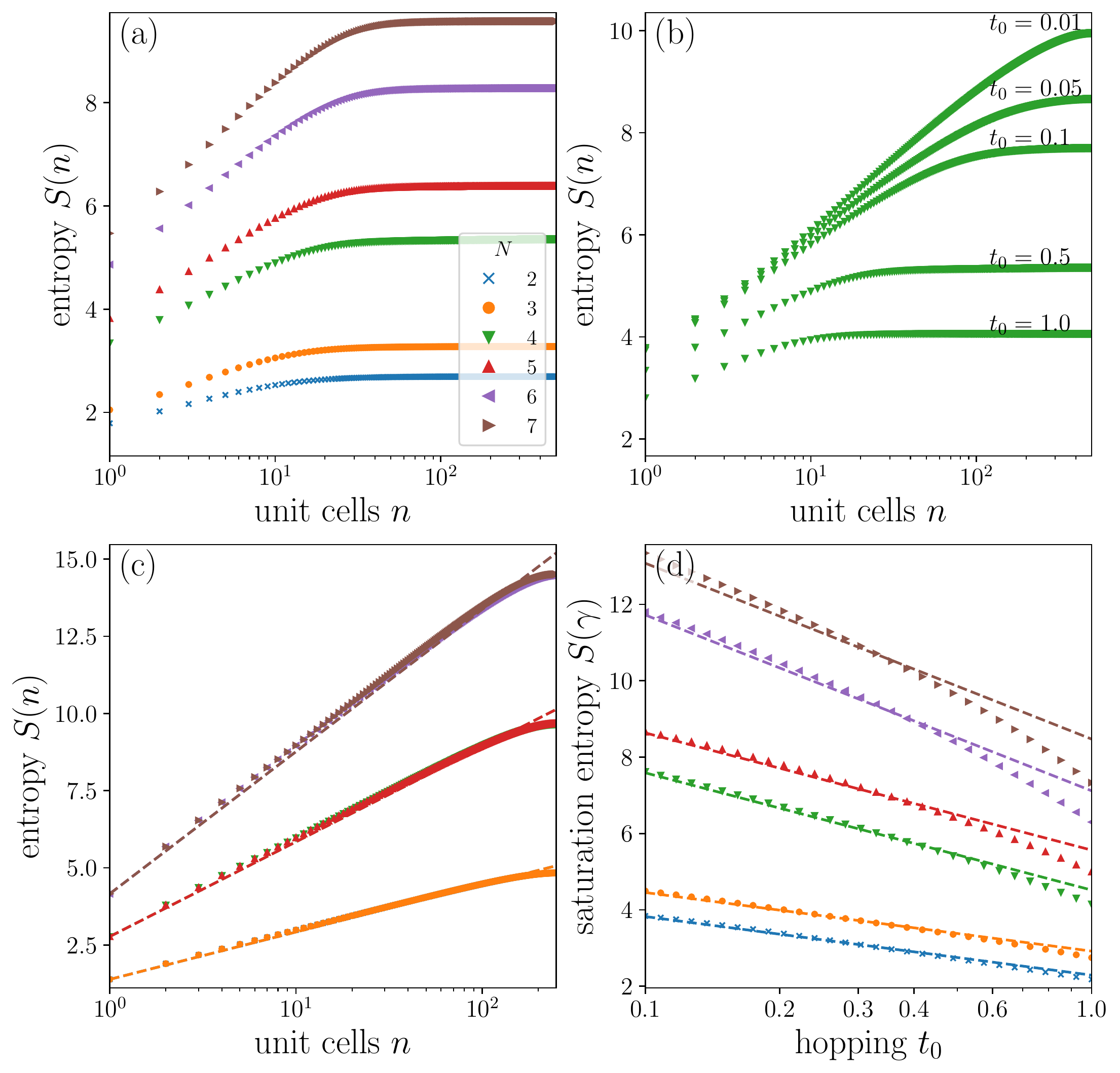}
\caption{(a) Entanglement entropy as a function of the subsystem size in unit cells $n$ at $t_0=0.5t$. The lattice models host an \ep{N}, with $N$ in between  2 and 7. (b) Typical evolution of the entropy when $N=4$, for different $t_0$ values.
(c) Entanglement entropy for $n \ll t/t_0$ (symbols) is fitted with Eq.~\eqref{s_square_short} (dashed lines) at $t_0=10^{-3}t$.
(d) Saturation entropy (symbols) as a function of $t_0$ for $n\gg t/t_0$ is fitted with Eq.~\eqref{s_square_large} (dashed lines).
The legend is shared for all panels and $t_0$ is in units of $t$.
}
\label{fig:ent_square}
\end{figure}

In this section, we examine the entanglement entropy of the ground state for Hamiltonians~\eqref{ham_square_tb} with fillings up to the EP energy. We present the results in Fig.~\ref{fig:ent_square}, which shows the entanglement entropy as a function of the number of cells in the subsystem, denoted as $n$. Panel (a) depicts a typical evolution of the entanglement entropy for various values of $N$. Similar to the diamond lattice case, the entanglement entropy displays a transition from logarithmic growth below the correlation length to a saturation behavior characteristic of an insulator above it. The crossover point shifts towards higher values for larger $N$ and smaller $t_0$ (as shown in Fig.~\ref{fig:ent_square}), which supports the notion that the correlation length varies as $\xi\sim N t/t_0$.

In the short-distance limit, for $t_0\ll t$, the entropy is represented in Fig.~\ref{fig:ent_square}(c) and is approximated by
\begin{equation}\label{s_square_short}
S(n\ll \xi) \approx \frac{2\lfloor N/2\rfloor}{3}\ln(8n).
\end{equation}
The results suggest that the central charge for the non-Hermitian Hamiltonians \eqref{ham_square_tb} is the same as that in the Hermitian system, i.e., $c=N/2$ for even $N$, and $(N-1)/2$ for odd $N$, where the factor of 2 in \eqref{s_square_short} 
arises due to the presence of two EPs in the Brillouin zone. 
This can be understood in the Hermitian limit $t_0=0$, where the models with odd $N$ consist of $N-1$ chains disconnected from each other, with each chain having entropy $S(n)\sim \frac{c}{3}\log(n)$ and $c=1$. The remaining central chain, responsible for the flat band, becomes a collection of isolated sites, and contributes nothing to $S$, as all entanglement among its sites is lost. 
For larger values of $t_0$, as seen in Fig.~\ref{fig:ent_square}(a), odd $N$ models start to deviate from the above-mentioned growth law.
At large distances $n\gg \xi$, the entropy eventually saturates and is approximated as
\begin{equation}\label{s_square_large}
S(n\gg \xi) \approx \frac{2\lfloor N/2\rfloor}{3}\ln(\xi)+\text{const.}.
\end{equation}
The numerical results presented in Fig.~\ref{fig:ent_square}(d) are well described by Eq.~\eqref{s_square_large}. However, it should be noted that this equation is only valid at smaller $t_0$, and as $t_0$ approaches the maximal reciprocal hopping amplitude $t$, the entropy in the large-$N$ models decreases faster than $\ln(\xi)$ towards zero. This can be attributed to the presence of a transverse skin effect induced by the unidirectional hopping $t_0$. As $t_0$ becomes stronger than the intrachain hopping, the entanglement entropy approaches zero, indicating that all the particles have moved to one side of the lattice.

\section{Comment on the relation between the two models}
\label{sec:comment}
In this section, we aim to investigate the connection between the two types of Hamiltonians. Specifically, we focus on the case of $N=3$ models, although our analysis applies equally to $N>3$. Although the Bloch Hamiltonians representing the two models yield different correlation functions, their counterparts in the continuum limit are identical, except for a rotation in the pseudo-spin space. Therefore, it is essential to understand the relationship between the correlation functions and their asymptotic behavior in the two models.

A unitary transformation $U=\exp(-i\pi S^y/2)$ transforms the Hamiltonian~\eqref{spin_ham} into the same basis as Hamiltonians~\eqref{ham_square_tb}, $\tilde h=UhU^\dag$,
\begin{equation}
\tilde h_k^{(1)} = -t\cos(k/2)S^z+\gamma S^x+i\gamma\sin(k/2)S^y.
\end{equation}
Here we introduce the superscript notation to distinguish the models in Sec.~\ref{sec:spin} from those in Sec.~\ref{sec:unidirectional} with $(1)$ and $(2)$, respectively.
The correlations in the transformed basis exhibit the same behavior for correlations, which at large distance still retain the anomalous power laws $1/n^2$ and $1/n^3$.
This is because in this new basis, the new correlators remain a linear combination of correlation functions found before, e.g.~Eqs~\eqref{corr_diamond_3_diag} and \eqref{corr_diamond_3_off}. 
The second Hamiltonian, lacking the alternation in gain-loss rates, reads
\begin{equation}
h_k^{(2)} = 2t\cos(k)S^z+\sqrt 2 t_0 S^x+i\sqrt2 t_0S^y,
\end{equation}
and has the power-law dependence $1/n^a$, $a\in\{2,3,4,5\}$.
Performing a similar rotation of $h_k^{(2)}$ and its eigenstates enables one to calculate the correlation function in the new basis and determine the power-law exponents $a\in\{2,3\}$.
This suggests that the two models are not directly related by a rotation, despite the second model being able to replicate the correlations of the first model. This can be attributed to the subdominance of the $a=4,5$ correlations in the linear combination of correlations. To understand this phenomenon, we study the low-energy continuum limit of the models near an EP.
Expanding $\tilde h^{(1)}_k$ to first order in momentum $k=\pi+p$ for small $p$ produces the Hamiltonian
\begin{equation}
\tilde h^{(1)}(p) = \frac{pt}{2}S^z + \gamma S^x+i\gamma S^y,
\end{equation}
which is identical to $h^{(2)}(p)$~\eqref{ham2_cont_3} under the identification $v=t/2$, $\Delta=\sqrt2\gamma$. 
Consequently, the correlation functions will acquire the power-law dependence $1/x^a$, $a\in\{2,3,4,5\}$.

A rotation back of $\tilde h^{(1)}(p)$ to $h^{(1)}(p)$ from Eq.~\eqref{spin_cont_3} reproduces the correlations from Eqs.~\eqref{diamond_cont_intra_3} and \eqref{diamond_cont_inter_3}.
This is possible since in the asymptotic limit, in the linear combination only the dominating term remains.
Performing the rotation $\tilde h^{(1)}(p)\to h^{(1)}(p)$ yields the following dependence in the correlations
\begin{eqnarray}
C^{(1)}_{1,1}(x)&\sim& 
\frac{\xi}{4 \pi x^2}
+\frac{i \xi^2}{4 \pi x^3}
-\frac{3 \xi^3}{8 \pi x^4}
-\frac{3 i \xi^4}{16 \pi x^5},
\notag\\
C^{(1)}_{2,2}(x)&\sim& 
-\frac{i \xi^2}{2 \pi x^3}-\frac{3 i \xi^4}{8 \pi x^5},
\notag\\
C^{(1)}_{3,3}(x)&\sim& 
-\frac{\xi}{4 \pi x^2}
+\frac{i \xi^2}{4 \pi x^3}
+\frac{3 \xi^3}{8 \pi x^4}
-\frac{3 i \xi^5}{16 \pi x^5},
\\
C^{(1)}_{1,2}(x)&\sim& 
-\frac{\sqrt2\xi}{8\pi x^2}
+\frac{i\sqrt2 \xi^2}{8 \pi x^3}
-\frac{3\sqrt2 \xi^3}{16 \pi x^4}
-\frac{3\sqrt2 i \xi^4}{16\pi x^5},
\notag\\
C^{(1)}_{1,3}(x)&\sim& 
-\frac{i \xi^2}{4 \pi x^3}
-\frac{3 i \xi^4}{16 \pi x^5},
\notag\\
C^{(1)}_{2,3}(x)&\sim& 
\frac{\sqrt2 \xi}{8 \pi x^2}
+\frac{i\sqrt2 \xi^2}{8\pi x^3}
+\frac{3\sqrt2 \xi^3}{16\pi x^4}
-\frac{3\sqrt2 i \xi^4}{16 \pi  x^5}.
\notag
\end{eqnarray}
If we keep only the dominant term in $1/x$ for each equation, we recover Eqs.~\eqref{diamond_cont_intra_3} and \eqref{diamond_cont_inter_3}. However, it is important to note that rotating back from Eqs.~\eqref{diamond_cont_intra_3} and \eqref{diamond_cont_inter_3} cannot reproduce the higher power terms $a=4,5$ in Eqs.~\eqref{ham2_cont_intra13} and \eqref{ham2_cont_inter13} for the model $h^{(1)}(p)$. This is because correlations computed for $h^{(1)}(p)$ must also include subdominant terms in $1/x$, as we saw earlier. We have checked that including terms up to $1/x^5$ reproduces the correlations $C^{(2)}_{i,j}(x)$ in the second class of models, as it should.

Our findings reveal that the two lattice Hamiltonians exhibit different power-law decay of correlations due different implementation of the non-Hermitian term, and they are not related by a unitary transformation. 
Therefore, the two models are distinct. However, in the continuum limit and to leading order in momentum near the exceptional point, the physics is identical, as they are related by a rotation in pseudo-spin space. 
To reveal this relation for correlation functions, one needs to go beyond the leading order in the asymptotic analysis of correlations in both models.

\section{Conclusions}
\label{sec:conclusions}
This study demonstrates the existence of exceptional points with arbitrary order $N$ in a \pt-symmetric diamond lattice implementation of a non-Hermitian Dirac Hamiltonian with general spin $S=(N-1)/2$. 
These EPs mark a transition from real to imaginary eigenvalues, which is reflected in the system's critical phase and the power-law decay of spatial correlations. The gain and loss rates in the lattice introduce a correlation length $\xi$, beyond which correlations are further suppressed with anomalous exponents of $1/x^2$ and $1/x^3$. Additionally, the lattice exhibits a charge-density accumulation at its edge due to a chemical potential difference between the edges. Non-zero charge currents in the ground state also accompany this effect.

The study also explores another class of systems built out of unidirectional Hatano-Nelson models instead of involving balanced gain and loss. 
Although both types of models are \pt\ symmetric, the system in Section~\ref{sec:spin} breaks time-reversal symmetry, while the models in Section~\ref{sec:unidirectional} obey both parity and time-reversal symmetry separately. 
These models exhibit power-law suppression of correlations, with an exponent proportional to the order of EP. The unidirectional hopping in the lattice creates a skin effect, which transfers charge to one side of the lattice stripe, resulting in a similar charge imbalance to the one observed in the previous lattice models. However, unlike the previous models, no charge currents were observed in the ground state.

Despite their differences, the models exhibit several similar characteristics. 
The effective low-energy models, developed near the exceptional point energy, are identical after a unitary transformation, and their correlation functions are related. Additionally, both models are subject to a particle-hole symmetry, which causes models with \ep{N} and $N$ odd to have a flat band. In the critical state, there is a general relationship between correlation functions $C_{i,j}$ for an empty and a filled flat band,
\begin{equation}\label{phs_N_conc}
C^{\nu=\frac{N+1}{2N}}_{i,j}(n) = (-1)^{i+j+1}C^{\nu=\frac{N-1}{2N}}_{N+1-j,N+1-i}(n),
\end{equation}
with filling $\nu=(N+1)/2N$ and $\nu=(N-1)/2N$ for an occupied and an empty flat band, respectively.
Moreover, both models display similar behavior in terms of entanglement entropy. 
The correlation length $\xi$ marks the transition from conventional critical behavior to insulating-like behavior, even if the spectrum has no gap. 
In the short-distance limit ($x \ll \xi$) that connects to the Hermitian limit as $\xi\to \infty$, the entropy grows logarithmically with the subsystem size. However, in the long-distance limit ($x \gg \xi$), the entropy saturates similarly to a gapped system and depends on the correlation length $\xi$.

\begin{acknowledgments}
The authors acknowledge illuminating discussions with F.~Pi\'echon regarding the electronic band structure singularities in Hermitian and non-Hermitian systems. 
This work was supported by a grant of the Romanian MCID, CNCS/CCCDI-UEFISCDI, under Project No.~PN-III-P4-ID-PCE-2020-0277, under the project for funding the excellence, 
Contract No.~29 PFE/30.12.2021, and ``Nucleu'' Program 27N/03.01.2023, Project No.~PN 23 24 01 04, and by the Ministry of Culture and Innovation and the National Research, Development and Innovation Office within the Quantum Information National Laboratory of Hungary (Grant No. 2022-2.1.1-NL-2022-00004) K134437, K142179.
\end{acknowledgments}

\onecolumngrid{}
\appendix

\section{Matrix representation for general spin operators}
\label{sec:spin_mat}
The spin operators in a basis $\{|S,m\rangle\}$ have the conventional matrix representation:
\begin{eqnarray}\label{mat_repr}
\avg{S,m|\hat S^x|S,m'} &=& \frac{1}{2}\sqrt{S(S+1)-mm'}(\delta_{m,m'+1}+\delta_{m,m'-1}),\notag\\
\avg{S,m|\hat S^y|S,m'} &=& \frac{i}{2}\sqrt{S(S+1)-mm'}(\delta_{m,m'+1}-\delta_{m,m'-1}),\\
\avg{S,m|\hat S^z|S,m'} &=& m\delta_{m,m'},\notag
\end{eqnarray}
with $m$ and $m'\in\{-S,-S+1,\ldots,S\}$, and $S$ denoting an arbitrary (half-)integer.

\section{Correlation functions}
\label{sec:correlations}
The central object of study are the correlation functions
\begin{equation}\label{corr_latt}
C_{i,j}(n) = \avg{c^\dag_{i,n} c_{j,0}^{}},
\end{equation}
which capture the correlations between sites $i$ and $j$ (with $i,j$ from 1 to $N$ internal degrees of freedom in the unit cell) at a distance of $n$ unit cells.
The expectation value $\avg{\dots}$ is taken with respect to the ground state of the non-Hermitian system.  For all our computations, we assume an electronic filling where all the energy levels up to the one associated with the exceptional point are occupied, and this exceptional point energy is typically set to zero. To obtain the expected value, we utilize the right-eigenvectors of the Hamiltonian.
The latter are not generally orthogonal to each other and therefore an orthogonalization is performed when there is more than a single occupied eigenstate~\cite{Herviou2019}. This procedure ensures that the trace implied when taking the expectation value over the occupied states is well-defined.
Since Hamiltonian right-eigenvectors are used in $\avg{\dots}$, the correlation matrix~\eqref{corr_latt}  $[C(n)]_{i,j}$, indexed by internal degrees of freedom $i,j$, is a Hermitian matrix.
It implies that in the tight-binding models considered in this work, for spinless fermions with $N$ sites in the unit cell, there are $N(N+1)/2$ distinct correlation functions to determine.
We referred throughout to the $N$ $C_{i,i}$ as the diagonal correlations and to the $N(N-1)/2$ $C_{i,j\neq i}$ as the off-diagonal correlations.

In continuum models, the convention is that correlation functions have a position $x$ argument (instead of unit cell $n$),
\begin{equation}\label{corr_cont}
C_{i,j}(x) = \avg{\psi^\dag_{i}(x) \psi_{j}^{}(0)},
\end{equation}
and the indices $i,j$ are flavors of the field operators $\psi_i(x)$.

\section{Particle-hole symmetry and the flat band}
\label{sec:phs_flat}
For odd $N$ models there is a zero-energy flat band crossing through the \ep{N}. 
For models filled to the EP, there are two relevant correlation functions, corresponding to the case of an empty or filled flat band, which amounts to a difference in the total filling of $1/N$. 

In the present case, the two sets of correlation functions are related by a particle-hole symmetry. 
That allows us to determine the correlation functions when including a filled flat band, from the usually simpler case, of an empty flat band.
The proof of the previous statement starts by noting that the lattice Hamiltonians in Eqs.~\eqref{ham_diamond} and~\eqref{ham_square_tb} are invariant under the particle-hole symmetry $\mc C$,
\begin{equation}\label{phs_latt}
\mc C c^\dag_{i,n}\mc C^{-1} = (-1)^{i}c_{N+1-i,-n}.
\end{equation}
This assumes an inversion center in the middle of the unit cell $n=0$, which is true for our \pt\ symmetric models.
The filling of the model without the flat band is $\nu=\frac{N-1}{2N}$, while, in the presence of a filled flat band, $\nu=\frac{N+1}{2N}$. 
The respective ground states at zero temperature are related under the transformation $\mc C$,
\begin{equation}\label{phs_GS}
\mc C\big|\nu=\frac{N+1}{2N}\big\rangle 
= \big|\nu=\frac{N-1}{2N}\big\rangle.
\end{equation}
They are both eigenstates of $H$ with the same energy, since the flat band is pinned at zero energy.
Then, using Eqs.~\eqref{phs_latt} and \eqref{phs_GS}, any correlation function for filling up to \ep{N}, including a filled flat band, is obtained from correlations with an empty flat band,
\begin{eqnarray}\label{phs_N}
C^{\nu=\frac{N+1}{2N}}_{i,j}(n) &=& (-1)^{i+j}\avg{c_{N+1-i,-n}^{}c^\dag_{N+1-j,0}}_{\nu=\frac{N-1}{2N}}\notag\\
&=&(-1)^{i+j+1}C^{\nu=\frac{N-1}{2N}}_{N+1-j,N+1-i}(n),
\end{eqnarray}
with the filling $\nu$ denoted explicitly, and the last equality follows under the translation symmetry.

In the continuum limit, the particle-hole transformation of the field operators reads,
\begin{equation}\label{phs_cont}
\mc C \psi_j^\dag(x)\mc C^{-1}=(-1)^j\psi_{N+1-j}(-x).
\end{equation}
It can be immediately seen that continuum Hamiltonians in the main text, $H=\sum_{i,j}\int dx \psi_{i}^\dag(x)h^{}_{ij}(x)\psi^{}_j(x)$, are also invariant with respect to it.

\section{Proof that the EPs are order \texorpdfstring{$N$}{N}}
\label{sec:epn_proof}
Here we prove that the Hamiltonian at the EP in the diamond lattice Eq.~\eqref{spin_cont_3} describes an \ep{N} by showing that there are $N$ degenerate eigenvectors at the EP.
For the choice $\mu=\gamma$, the Hamiltonian at $p=0$ reads
\begin{equation}
h(0)=i\gamma S^y + \gamma S^z,
\end{equation}
with a single zero-energy eigenvalue and algebraic multiplicity $N$.
A unitary transformation does not change the order of the EP. Performing such a rotation with an angle $\pi/2$ around the $y$ axis gives
\begin{equation}
\tilde h(0) = \gamma S^+,\quad S^+=S^x+iS^y.
\end{equation}
To prove that the geometric multiplicity is 1, i.e.,~there are $N$ degenerate eigenvectors, we demonstrate that the Hamiltonian or the ladder operator $S^+$ is nilpotent of index $N$. 
Using Eqs.~\eqref{mat_repr} for $S^x$ and $S^y$, it follows readily that the matrix elements of powers of $S^+$ read 
\begin{equation}\label{sn}
\avg{m|(S^+)^n|m'} = \prod_{i=0}^{n-1}
\sqrt{S(S+1)-(m+i)(m+i+1)}\delta_{m,m'-n}.
\end{equation}
There are two cases to consider for $n$. If $n=N$, then $\delta_{m,m'-N}=0$ for $m$ and $m'\in\{-S,-S+1,\ldots,S\}$, and
\begin{equation}
(S^+)^N=0.
\end{equation}
If $1\leq n<N$, then there is at least one matrix element that is nonzero with $m=-S$. For this element, back in Eq.~\eqref{sn}, it follows readily that
\begin{equation}
\sqrt{S(S+1)-(-S+i)(-S+i+1)}>0, \text{ for any }i<N-1.
\end{equation} 
Therefore,
\begin{equation}
(S^+)^n\neq 0 \text{ for } n<N,
\end{equation}
which completes the proof that $S^+$ is nilpotent of index $N$.
Hence, $S^+$, and, by extension $h(0)$, describe an EP of order $N$~\cite{Kato1995,Wiersig2022}.
The same analysis holds for the choice $\mu=-\gamma$, where $S^-$ is nilpotent of index $N$.

\section{Illustrative calculation of correlations}
\label{sec:example}

This appendix provides examples for calculating correlation functions, which are representative of the non-Hermitian setup. We have selected a few examples that highlight the unique features of this setup. Specifically, we will derive a correlation function for the $N=2$ continuum model~\eqref{continuum_h}, the $N=3$ continuum model~\eqref{spin_cont_3}, and a more complex example for the $N=3$ lattice model~\eqref{ham_square_tb}. This latter example also demonstrates how particle-hole symmetry can be used to determine correlation functions at $\nu=2/3$ based on those at $\nu=1/3$. The remaining correlation functions in the main text can be straightforwardly determined using the same approach.

\textbf{Example 1.}
The Hamiltonian~\eqref{continuum_h} for $N=2$ has the following eigenvalues and eigenstates:
\begin{alignat}{2}
E_1 &= vp, & \phi_1 &= (1,0)^T,\notag\\
E_2 &=-vp,\quad & \phi_2 &= \frac{(-\Delta\text{sgn}(p),2v|p|)^T}{\sqrt{\Delta^2+4v^2p^2}},
\end{alignat}
with $T$ denoting transpose.
As an example, we compute the off-diagonal propagator $C_{1,2}(x)=\avg{\psi_1^\dag(x)\psi_2^{}(0)}$, which exists only in the non-Hermitian case, due to the coupling $\Delta$ between left and right movers,
\begin{equation}\label{c12_ep2}
C_{1,2}(x) = - \frac{1}{2\pi}\int_0^{\infty}dp e^{-ipx} \frac{2vp\Delta}{\Delta^2+4v^2p^2}.
\end{equation}
The integral is solved by ensuring its convergence with a cutoff $\delta$, $x\to x-i\delta$, which is safely set to zero in the final result~\cite{Giamarchi2004}.
In the large-distance limit $x\gg v/\Delta$, the integral behaves as
\begin{equation}
C_{1,2}(x) \sim \frac{v}{\pi \Delta x^2}.
\end{equation}
In the short-distance limit $x\ll v/\Delta$, integration by parts in Eq.~\eqref{c12_ep2} allows one to single out the diverging contribution at the origin, yielding
\begin{equation}
C_{1,2}(x) \sim \frac{\Delta}{4\pi v}\ln(\frac{\Delta x}{2v})+\frac{i\Delta}{8v}
+\frac{\gamma\Delta}{4\pi v} 
\sim \frac{\Delta}{4\pi v}\ln(\frac{\Delta x}{2v}),
\end{equation}
where the first two contributions come from the boundary term, and we neglected terms $\mc O(x)$, and $\gamma$ is the Euler-Mascheroni constant. 
In the final result we keep the dominating contribution at $x\ll v/\Delta$.
The vanishing right-left correlators in the Hermitian limit are obtained as the non-Hermitian coupling is $\Delta\to 0$.

\textbf{Example 2.} Let us consider the $N=3$ continuum model~\eqref{spin_cont_3} at $\nu=1/3$. 
Since the flat band is empty, in order to determine the correlations, it is sufficient to know the occupied eigenstates and their corresponding eigenvalues,
\begin{alignat}{2}
E_1 &= pt/2,\quad & \phi_1 &= \frac{1}{2}(1,-\sqrt2,1)^T,\notag\\
E_3 &= -pt/2,\quad &\phi_3 &= \frac{1}{2}(1, \sqrt2,1)^T
- \frac{2pt\gamma}{p^2t^2+\gamma^2}(1,\frac{2\sqrt2 \gamma}{pt}, -1)^{T}.
\end{alignat}
As an example, we look at the correlator
\begin{equation}
C_{1,2}(x) = -\frac{\sqrt 2}{8\pi}\int_{-\infty}^{0} dp e^{-ipx} 
+ \frac{\sqrt2}{8\pi}\int_0^{\infty}dp e^{-ipx} \frac{(pt-2\gamma)^2(p^2t^2-4\gamma^2)}{(p^2t^2+4\gamma^2)^2}.
\end{equation}
The divergent integrals are regularized with a positive cutoff $\delta$, $x\to x+i\delta$, for $p<0$, and $x\to x-i\delta$, for $p>0$.
In the large distance limit, $x\gg t/\gamma$, and taking the limit $\delta\to 0$ gives the first integral $-i\sqrt2/8\pi x$. The second integral is expanded to next orders in $1/x$ yielding
\begin{equation}
\frac{i\sqrt2}{8\pi} \bigg[\frac{1}{x}+\frac{it}{\gamma x^2}+\frac{t^2}{\gamma^2x^3}+\ldots\bigg].
\end{equation}
The first term cancels between the two integrals, and keeping the dominant term in the large distance limit yields the result in the main text, with $\xi=t/\gamma$,
\begin{equation}
C_{1,2}(x\gg \xi) \sim -\frac{\sqrt 2\xi }{8\pi x^2}.
\end{equation}
In the short distance limit, $x\ll \xi$, one takes $\delta\to 0$, $x>\delta$. The first integral is divergent and with the same expression as above.
The second one is solved in the limit of small $\delta$, and the same dominant behavior is found as in the first integral. Adding the two terms yields
\begin{equation}
C_{1,2}(x\ll\xi) \sim -\frac{i\sqrt2}{4\pi x},
\end{equation}
which is usual for a one-dimensional Hermitian system, but here it happens only below the correlation length.

\textbf{Example 3.} Let us consider the lattice model~\eqref{ham_square_tb} for $N=3$.
The three eigenvalues and eigenstates of the lattice Hamiltonian are, respectively,
\begin{alignat}{2}
E_1 &= -2t_1\cos(k),\quad & \phi_1 &= (1,0,0)^T,\notag\\
E_2 &= 0, & \phi_2 &= \frac{(-t_0\text{sgn}(\cos(k)), 2t_1|\cos(k)|,0)^T}
{\sqrt{t_0^2+4t_1^2\cos(k)^2}}, \\
E_3 &= 2t_1\cos(k),\quad &\phi_3 &= \frac{(t_0^2,-4t_0t_1\cos(k),8t_1^2\cos(k)^2)^T}{t_0^2+8t_1^2\cos(k)^2}.
\end{alignat}
First, at $\nu=1/3$ filling, we consider $C^{\nu=1/3}_{1,1}(n)=\avg{c_{1,n}^\dag c_{1,0}^{}}$,
\begin{equation}
C^{\nu=1/3}_{1,1}(n)
=\frac{1}{2\pi}\int_0^{\pi/2}dk e^{-ikn}
+\frac{1}{2\pi}\int_{\pi/2}^\pi dk e^{-ikn} \frac{t_0^4}{(t_0^2+8t_1^2\cos(k)^2)^2}+(n\to -n).
\end{equation}
The first integral is immediate and, together with $(n\to -n)$ contribution, yields $\sin(n\pi/2)/n$. The second one can be solved asymptotically by first moving the contour of integration in the complex plane on the path $\{\pi/2, \pi/2-is, \pi-is,\pi\}$, where the path is straight between the vertices denoted before. 
The parameter $s$ is real and is sent to infinity. Then taking the limit of large distance $n$, one expands to first orders in $1/n$. 
The $1/n$ term in the second integral cancels exactly the first integral, leaving the dominating behavior
\begin{equation}\label{11_1over3}
C^{\nu=1/3}_{1,1}(n) \sim 
-\frac{32t_1^2}{\pi t_0^2}\frac{\sin(n\pi/2)}{n^3}.
\end{equation}  

An example of a lattice correlator at $\nu=2/3$ filling shows some distinct features of the non-Hermitian problem.
To take expectation values or determine correlations which involve taking a trace over all states, it is necessary to orthogonalize the eigenstates in the Fermi sea. 
This can still be performed analytically, since for the $N=3$ case there are at most two occupied eigenstates at zero temperature.
The orthogonalization procedure gives the occupied eigenvectors,
\begin{alignat}{2}
\phi_1 &= (1,0,0)^T, \quad\tilde\phi_2 = (0,1,0)^T,\quad & k&\in(-\frac\pi 2,\frac\pi 2),\notag\\
\phi_2 &=\frac{(t_0,-2t_1\cos(k),0)^T}{\sqrt{t_0^2+4t_1^2\cos(k)^2}},\quad
\tilde\phi_3 = \frac{(2t_0^2t_1\cos(k),t_0^3, -4t_1\cos(k)(t_0^2+4t_1^2\cos(k)^2))^T}
{\sqrt{t_0^2+4t_1^2\cos(k)^2}(t_0^2+8t_1^2\cos(k)^2)},\quad
& k&\in [-\pi,-\frac\pi 2)\cup (\frac\pi2,\pi],
\end{alignat}
where tilde denotes that the respective eigenvector was modified to become orthogonal to the other one. 
Then, an example of a simple correlation functions is
\begin{equation}
C^{\nu=2/3}_{3,3}(n) = \frac{1}{2\pi}\int_{\pi/2}^{\pi}dke^{-ikn}
\frac{16t_1^2\cos(k)^2(t_0^2+4t_1^2\cos(k)^2)}{(t_0^2+8t_1^2\cos(k)^2)^2} + (n\to -n).
\end{equation}
Moving the integration contour in the complex plane on the path $\{\pi/2, \pi/2-is, \pi-is,\pi\}$ as $s\to \infty$, and taking the limit of large $n$, gives the result
\begin{equation}
C^{\nu=2/3}_{3,3}(n) \sim \frac{16it_1^2}{\pi t_0^2 n^3}e^{-i\pi n/2} + (n\to-n)
=\frac{32t_1^2}{\pi t_0^2} \frac{\sin(n\pi/2)}{n^3}.
\end{equation}
For the $N=3$ model, such a cumbersome calculation is shortened by using the particle-hole symmetry~\eqref{phs_latt} and the previously determined correlation function at $\nu=1/3$ in Eq.~\eqref{11_1over3},
\begin{equation}
C^{\nu=2/3}_{3,3}(n) = (-1)^{7} C^{\nu=1/3}_{1,1}(n) = \frac{32t_1^2}{\pi t_0^2} \frac{\sin(n\pi/2)}{n^3}.
\end{equation}

\twocolumngrid

\bibliographystyle{apsrev4-2}
\bibliography{bibl}
\end{document}